\documentclass[aps,groupedaddress,nofootinbib]{revtex4}

\usepackage[bookmarks=true,bookmarksopen=true,pdfhighlight=/I,pdfpagemode=UseOutlines]{hyperref}
\usepackage[latin1]{inputenc}
\usepackage{graphicx}
\usepackage{amssymb,amsmath}
\usepackage[usenames,dvipsnames]{color}
\usepackage{hyperref}

\newcommand{\be}{\begin{equation}}
\newcommand{\ee}{\end{equation}}
\newcommand{\bea}{\begin{eqnarray}}
\newcommand{\eea}{\end{eqnarray}}	
\newcommand{\nn}{\nonumber\\}
\newcommand{\ba}{\begin{array}}
\newcommand{\ea}{\end{array}}

\newcommand{\w}{\omega}

\newcommand{\rfv}{|0\rangle_f}
\newcommand{\lfv}{_f\langle 0 |}
\newcommand{\rmv}{| 0 \rangle}
\newcommand{\lmv}{\langle 0 |}

\newcommand{\vp}{\vec{p}}
\newcommand{\vk}{\vec{k}}
\newcommand{\vq}{\vec{q}}
\newcommand{\vx}{\vec{x}}
\newcommand{\e}{\eta}

\newcommand{\lag}{\mathcal L}

\newcommand{\st}{\sin \theta}
\newcommand{\sst}{\sin^2 \theta}

\newcommand{\ct}{\cos \theta}

\definecolor{MyDarkGray}{RGB}{140,140,140}

\begin{document}

\title{D-Foam Induced Flavour-Condensates and Breaking of Supersymmetry \\  in Free Wess-Zumino Fluids}


\author{Nick E. Mavromatos}
\affiliation{CERN, Theory Division, CH-1211  Geneva 23, Switzerland; \\On leave from: King's College London, Department of Physics, Strand, London WC2R 2LS, UK.}

\author{Sarben Sarkar and Walter Tarantino}
\affiliation{King's College London, Department of Physics, Strand, London WC2R 2LS, UK.}



\begin{abstract}
Recently \cite{mavrosarkar,mst1}, we argued that a particular model of
string-inspired quantum space-time foam (D-foam) may induce oscillations and mixing
among flavoured particles. As a result, rather than the mass-eigenstate vacuum,  the correct ground state to describe the underlying dynamics is the \emph{flavour vacuum}, proposed some time ago by Blasone and Vitiello as a description of quantum field theories with mixing.
At the microscopic level, the breaking of target-space supersymmetry is induced in our space-time foam model by the relative transverse motion of brane defects. Motivated by these results, we show that the flavour vacuum, introduced through an inequivalent representation of
the canonical (anti-)~commutation relations, provides a vehicle
for the breaking of supersymmetry (SUSY) at a low-energy effective field theory level; on considering the flavour-vacuum expectation value of the energy-momentum tensor and comparing with the form of a perfect relativistic fluid, it is found that the bosonic sector contributes
as dark energy while the fermion
contribution is like dust. This indicates a strong and novel
breaking of SUSY, of a non-perturbative nature, which may characterize the low energy field theory of certain quantum gravity models.
\end{abstract}

\maketitle

\section{Introduction}

We have, in an earlier work \cite{mavrosarkar,mst1}, investigated
the behaviour of an effective vacuum condensate in a field theory arising from a brane model of space-time foam (\textit{D-Particle
Foam Model}). Within a Brane-World context (\emph{cf.} 
\cite{mavrosarkar} and citations therein), this condensate is consistent with a flavour vacuum \cite{blasone} of a string model (in a low-energy limit) that includes string excitations corresponding to particles of different
flavours (on a D3 brane Universe) and a foam of D-particles (in the bulk); such a
condensate was described through the formalism of Quantum Field Theory (QFT) on a curved (classical)
background \cite{birrell} and the Blasone-Vitiello (BV) formalism for \textit{flavour
mixing} \cite{blasone}.

In our earliest investigation \cite{mavrosarkar}, a theory of two real scalar fields with flavour mixing,
on a specific expanding 1+1 dimensional universe, was considered. The equation of state $p=w\rho$ (with $p$ pressure and $\rho$ energy density),  associated with the
flavour vacuum was found to have $w=-1$  (on using a suitable regularisation).
In \cite{mst1} we claimed that such a behaviour would characterize also the 3+1 dimensional case, in a generic Friedmann-Robertson-Walker (FRW)  universe.
Moreover, we analyzed a theory with two Dirac fields (complex spin 1/2) on a 3+1 dimensional FRW universe. In this context,
the $w$ for the equation of state of the flavour vacuum was found to be in the range $-1/3<w<0$ (the precise value being determined by the cut-off for the momenta imposed
on the QFT theory), leading to a contracting universe.  Therefore, it has been advocated that this vacuum
condensate in a realistic theory might contribute to the Dark Energy fluid that
has been suggested as an explanation of the current acceleration
of the Universe. This is in agreement with the general arguments of ref.~\cite{capocosmo}, except that the analysis of the model of \cite{mavrosarkar,mst1} was done more consistently in a curved background, entailing particle creation and a dynamically generated cut-off scale. However the bosonic and fermionic particles were not considered as part of the same supersymmetric multiplet.
Hence for an observer on the D3 brane a detailed signature of any SUSY breaking needs to be sought. This will be effected by considering properties of a flavour vacuum state in a supersymmetric context.

In our microscopic brane model of \cite{mavrosarkar,mst1}, the seeds of microscopic Lorentz violation are provided by the recoil of D-particles, interacting with flavour matter states. An effective low energy QFT description of this stringy picture needs to reflect this. Hence it is important to note at this point that a QFT, based on the flavour Fock space of the BV formalism, is necessarily Lorentz violating.   As remarked in \cite{mst1}, this
distinct behaviour in the collective properties between fermions and bosons is a novel non-perturbative measure
of Supersymmetry (SUSY) breaking, which is consistent with the
Lorentz-violating nature of the BV formalism.

In this article we embark on a detailed investigation of this point, considering as a simple, but instructive, example a \emph{supersymmetric fluid} free Wess-Zumino (WZ) model with flavour
mixing.~\footnote{A work  \cite{capolupo} has appeared contemporaneously with the first draft of this article, available on the archives,
where a free flavoured WZ model has also been analysed from the point of view  of SUSY breaking by the flavour condensate. However, the authors of that work,
in contrast to our analysis here, examined only the energy part of the stress tensor (with results similar to ours) but did not analyze the pressure part of the fluid
and thus the relevant equation of state. The latter requires a technically much more involved and highly non-trivial operator analysis, and, in fact, is directly relevant for
a proper classification of the type of SUSY breaking involved, as well as the phenomenology and, in particular, the cosmology of such models, in which context their true significance
appears. Moreover, in view of our motivation outlined in section \ref{sec:dfoam}, the philosophy of our approach and the reasoning for the selection of the flavour vacuum are entirely
different.
This further differentiates our approach from theirs.} The structure of the article is as follows:
in section \ref{sec:dfoam} we introduce the reader to the concept of the medium of the D-particle space-time foam,
which constitutes a background space-time whose interactions with open-string flavoured excitations in the Brane world necessitate
the introduction of the flavour Fock states as the physically correct excitation spectrum. This illustrates that the interactions of a model can pick up the flavour vacuum as the physically correct vacuum, despite the fact that the latter may not be the minimal energy state in an effective theory. In the BV formalism this state has been associated with the vacuum state associated with the Fock space of an inequivalent representation of the field canonical commutation relations.

In this work, the two flavour generalisation of the free supersymmetric Wess-Zumino (WZ) model,  will be viewed as an ``effective''
field theory of ``free'' fields, however now viewed as (Fock) excitations of the ``flavoured'' vacuum. In Section \ref{sec:dfoam} we motivate 
the use of the flavoured supersymmetric WZ model as a prototype effective low-energy field theory, describing flavour mixing induced by quantum-gravity space-time foam, in particular 
stringy D-particle foam~\cite{Dfoam}. In Section \ref{setup} the model is introduced, with flavour mixing implemented through the BV formalism.
In Section \ref{features} we outline the steps of the calculations of the expectation value of the energy momentum tensor in the flavour vacuum. Further calculational details appear in subsection \ref{calculations}, and we present our results and comments in subsections \ref{results} and \ref{sec:eos}. Not only a positive flavour vacuum energy is demonstrated, but the equations of state for the fermion and boson parts are different, the bosons resembling a dark energy fluid (of cosmological type), while the fermions behave like dust. This indicates a strong breaking of SUSY, of a novel type, going beyond the conventional supersymmetry breaking situations characterised only by a non-zero vacuum energy.

Other scenaria, following from non-standard model physics, which may lead to flavour condensates (and possible SUSY breaking) will be considered in a forthcoming publication~\cite{new}.

\section{Motivation: D-particle Foam as a medium necessitating Flavour Fock States}\label{sec:dfoam}

\subsection{Features of the Model}

\begin{figure}[ht]
\centering
\includegraphics[width=9cm]{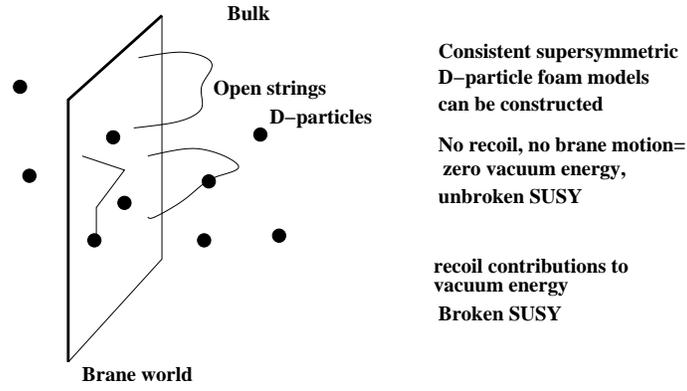} \vfill \vspace{2cm}
\includegraphics[width=9cm]{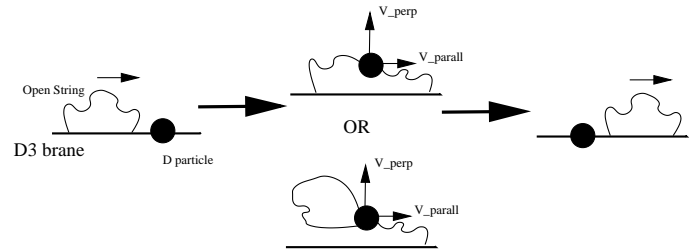}
\caption{\it \underline{Upper}: schematic
representation of a generic D-particle space-time foam model, in which matter particles
are treated as open strings propagating on a D3-brane, and the higher-dimensional
bulk space-time is punctured by D-particle defects. \underline{Lower}: details of the process
whereby an open string state propagating on the D3-brane is captured by a D-particle
defect, which then recoils.
This process involves an intermediate composite state that persists for a period
$\delta t \sim \alpha ' E$, where $E$ is the energy
of the incident string state, which distorts the surrounding
space time during the scattering, leading to an effective refractive index
but \emph{not} birefringence.The components of the recoil velocity perpendicular to the D3 brane world lead to vacuum energy contributions and thus target-space supersymmetry breaking.}%
\label{fig:recoil}
\end{figure}

The supersymmetric (string) model illustrated in the upper
panel of Fig.~\ref{fig:recoil}~\cite{Dfoam} will serve as a concrete framework for D-foam phenomenology.
In it, our Universe, after appropriate compactification, is represented  as a Dirichlet three-brane (D3-brane), propagating in a 10-dimensional bulk
space-time punctured by D-particle defects\footnote{Since an isolated D-particle cannot exist~\cite{polchinski},
because of string gauge flux conservation, the presence of a D-brane is essential.}.
As the D3-brane world moves
through the bulk, the D-particles cross it. To an observer on the D3-brane the model looks
like `space-time foam' with defects `flashing' on and off as the D-particles cross it: `D-foam'
denotes this structure. {\it If no relative
motion of branes takes place {in the bulk}} target-space supersymmetry,
implies the vanishing of the ground-state energy of the configuration.
However, if there is relative motion in the bulk, SUSY is broken and there are non-trivial forces among the D-particles as well as between the D-particles, the brane world  and orientifolds~\cite{Dfoam}. The resulting non-zero contribution to the energy is proportional to $v^2$ for \emph{transverse} relative motion of branes with different dimensionalities, and to $v^4$ for branes of the same dimensionality
(there is \emph{no} contribution to the energy of a p-brane-world from motion of other branes in directions \emph{parallel} to its longitudinal directions). There is also a dependence on the relative distance of the various branes. We shall now give a quite heuristic analysis of the effects of a gas of D-particles on the flavoured string state on the brane world. (A more string theoretic calculation will be given later which will model more rigorously the flavour changing feaures of the interaction of D-particles.) In particular, the interaction of a single D-particle, that lies far away from the D3 brane (i.e. a compactified D8) world, and moves adiabatically with a
small velocity $v_\perp$ in a direction transverse to the brane, results in the following potential~\footnote{For brevity, in what follows we ignore potential contributions  induced
by compactification of the D8 brane worlds to D3 branes, stating only the expressions for the induced potential on the uncompactified brane world as a result of a stretched string between 
the latter and the D-particle - the compactitication does not affect our arguments on the negative energy contributions to the brane vacuum energy.}~\cite{Dfoam}
\begin{eqnarray}
\label{D0-D8-long}
\mathcal{V}_{D0-D8}^{long} =  + \frac{r\,(v^{\rm long}_\perp)^2}{8\pi\alpha^{\prime}}~, ~r \gg \sqrt{\alpha^{\prime}}.
\end{eqnarray}
On the other hand, a D-particle close to the D3-brane (compactified D8), at a distance $r' \ll \sqrt{\alpha '} $,  moving adiabatically in the perpendicular direction
with a velocity $v^{\rm short}_\perp$
will induce the following potential to it:
\be\label{pot1}
\mathcal{V}_{D0-D8}^{\rm short}=  - \frac{\pi\alpha^\prime (v_\perp^{\rm short})^2}{12{r'}^3}.
\ee

D-foam involves different configurations of D-particles, and so one has to average
over appropriate populations and quantum fluctuations~\cite{mst1,sarkar} (due to recoiling D particles with (nine-dimensional) recoil velocities $v_i$). Recoil fluctuations \emph{along the D3 brane universe} represent a stochastic fraction of the incident momentum of the D-particle~\cite{sarkar}. Due to the D-particle fluctuations in the transverse direction of the brane worlds, there would also be contributions to the potential energy of the brane (\emph{cf.} (\ref{D0-D8-long}), (\ref{pot1}), where now the quantities involving $v^{\rm short, long}_\perp$ should be averaged over populations of D-particles and quantum fluctuations). We can plausibly use a general parametrization of the foam fluctuations as follows:
\begin{eqnarray}\label{foam}
&& \delta v_A = g_s \frac{r_\parallel}{M_s} p_A~, A=1,2,3 \quad \Rightarrow \quad \langle r_\parallel \rangle = 0, ~ \langle r_\parallel^2 \rangle = \sigma_\parallel ^2 \ne 0~\nonumber \\
&& v^{\rm short}_\perp \equiv v^{\rm short}_\alpha~, \alpha =4, \dots 9  \quad \Rightarrow \quad \langle v^{\rm short}_\alpha  \rangle = 0, ~ \langle v^{\rm short}_\alpha v^{\rm short}_\beta  \rangle = \delta_{\alpha\beta} {\sigma_{\rm short}'}^2 \ne 0~, \alpha, \beta = 4 \dots 9~, \nonumber \\
&& v^{\rm short}_\perp \equiv v^{\rm long}_\alpha~, \alpha =4, \dots 9  \quad \Rightarrow \quad \langle v^{\rm long}_\alpha  \rangle = 0, ~ \langle v^{\rm long}_\alpha v^{\rm long}_\beta  \rangle = \delta_{\alpha\beta} {\sigma_{\rm long}'}^2 \ne 0~, \alpha, \beta = 4 \dots 9~,
\nonumber \\
&& \langle v^{\rm short,long}_\alpha v^{\rm short, long}_j  \rangle = 0~.
\end{eqnarray}
Our
stochastic foam model has vanishing correlators of odd powers of $v_i$ of the recoil velocity, but non-trivial correlators of even powers. Above, $\langle \dots \rangle $ indicates averaging over both quantum fluctuations and the ensemble of D-particles in the foam; $A=1,2,3$ denote the longitudinal dimensions of the D3-brane world. The  $\delta v_A$, $A=1,2,3$, represent the recoil velocity components of the D-particle during scattering. In the absence of any interaction with matter strings there is still a velocity $v_i$ expressing the quantum fluctuations of individual D-particles.

 We consider the process of capture by a quantum-fluctuating  (heavy, non-relativistic) D-particle of an open string representing a neutral flavoured excitation, such as a neutrino (\emph{cf.}  lower figure in fig.~\ref{fig:recoil}). By momentum conservation during this scattering event one has
\begin{eqnarray}\label{foam2}
&& \langle (p_1 + p_2)_A \rangle = \frac{M_s}{g_s} \langle \delta v_A \rangle = 0~, \nonumber \\
&& \frac{M_s}{2g_s} \langle (v_A)_1^2 \rangle + E_1 = \frac{M_s}{2g_s} \langle (v_A + \delta v_A)_1^2 \rangle + E_2 + \delta E~, \quad E_I = \sqrt{p_A^2 + m_I^2}~, \quad I = 1,2~.
\end{eqnarray}
We have assumed that \textit{kinematical} constraints can be estimated by treating the open string states as relativistic free fields of typical mass $m_I$, and  the D-particles as
heavy (non-relativistic)particles, with masses $M_s/g_s \gg m_I$, where $M_s$ is the string scale, and $g_s< 1$ is the (weak) string coupling. $\delta E$ denotes \emph{brane energy fluctuations} in the transverse direction to the brane, which in view of (\ref{D0-D8-long}), (\ref{pot1})  (with $v_{D0} = v^{\rm short, long}_\perp$), are due both to long and short-distance D-particles. In our dilute-foam approximation we may take:
\begin{equation}\label{energyflct}
\delta E = \delta E_{\rm short} (\sigma_{\rm short}'(p)) + \delta E_{\rm long}(\sigma_{\rm long}')~,
\end{equation}
where according to our previous discussion only the short-distance stochastic-foam fluctuations may depend on the (average) momentum $p$ of the propagating matter state. In view of the $v_\perp^2$ dependence of the respective fluctuations (\emph{cf.} (\ref{D0-D8-long}), (\ref{pot1}))
both types of energy fluctuations are proportional to the respective velocity fluctuations $\sigma_\perp'$.

Averaging over foam fluctuations, and
Eq. (\ref{foam2}) implies that the (\emph{average}) energy $E_2$ of a string excitation, corresponding to a state of mass $m_1$,  propagating in the foam, will be modified from its free
value $E_1 = \sqrt{p^2 + m_1}$ by:
\begin{equation}\label{energycons}
E_2 = \sqrt{p^2 + m_1} -  \frac{g_s}{2 M_s}\sigma_\parallel^2 \, p^2 + \delta E_{\rm short}(\sigma_{\rm short}'(p)) + \delta E_{\rm long}(\sigma_{\rm long}')~.
\end{equation}
Since the signature of the quantum energy fluctuations is \emph{not fixed}, all kinds of situations for the foam dressed masses are allowed
in such processes. For instance, in case of flavoured states, where in the mass eigenstate spectrum, both masses are allowed,
\emph{i.e.} one may have $m_1 >  m_2$, when $\delta E < 0 $ is predominantly due to short range contributions (\ref{pot1}), while $m_1 > m_2$ in case
$\delta E$ is dominated by the long range contributions. On the other hand, when the massive state is not flavoured, then necessarily $m_1=m_2$
which is either achieved by the equality $ -  \frac{g_s}{2 M_s}\sigma_\parallel^2 \, p^2 + \delta E_{\rm short}(\sigma_{\rm short}'(p)) + \delta E_{\rm long}(\sigma_{\rm long}' )= 0$, or by means of appropriately modified dispersion relations, and hence a vacuum refractive index proportional to the foam fluctuations. Notice that these selection rules are allowed, as they express a link between the transverse foam fluctuations to the longitudinal ones, which are free parameters of the model. The net $\delta E$ can be both positive and negative, which would be compatible with oscillations. 

A simple illustrative analysis can be effected by using a piecewise continuous distribution $\textsc{p}(\eta)$ of D-particles in the (transverse or bulk direction) $\eta$ parametrised as
\[\textsc{p}(\eta)=\left\{
                \begin{array}{cc}
                  \frac{{\eta}^{3}}{L^{4}}, & \mbox{when $ 0<{\eta} <a$} \\ & \\
                 \frac{b}{L}, & \mbox{when $ a<{\eta}<L-a$} \\ & \\
                  -\frac{(\eta-L)^{3}}{L^{4}}, & \mbox{ $ L-a<{\eta} <L$} \\
                \end{array}\right.\]
(We have suppressed a normalisation factor $\textsl{N}$ which is given by $\textsl{N}=(b+\frac{a^{4}}{2L^{4}}-\frac{2ab}{L})^{-1}$.) The total energy 
$\cal{E}$ exerted on the D-brane by the D-particles ${\cal E} \propto \int_0^L d\eta \, \textsc{p}(\eta)\,  \tilde{\varepsilon}(\eta) $, with 
$ \tilde{\varepsilon}(\eta)$ denoting the linear bulk energy 
density at position $\eta$, 
satisfies (upon using the structures (\ref{D0-D8-long}), (\ref{pot1}) for $\tilde{\varepsilon}(\eta)$ appropriately):
\begin{equation}\label{energy}
    {\cal E} \propto {-[a v_{L}(b+\frac{a^{4}}{5L^{4}})+\frac{{\alpha '}^{2}a}{L^{4}}v_{S}]+\frac{v_{L}}{2}(b L+\frac{a^{4}}{4 L^{3}})}
\end{equation}
where
\begin{equation}\label{zero}
    {v_{L}}=\frac{(v^{\rm long}_\perp)^{2}}{8\pi}, \; {v_{S}}=\frac{\pi (v^{\rm short}_\perp)^{2}}{12}.
\end{equation}
The vanishing of $\cal E$ leads to the condition
\begin{equation}
    b \sim {\frac{{\alpha '}^{2} a}{L^{4}}} \frac{v_{S}}{{v_{L}}(\frac{L}{2}-a)}
\end{equation}
provided $a \ll {\alpha '}^{\frac{1}{2}}$, and so $\textsc{p}(\eta)$ displays a sub-stringy structure (consistent with the validity of the limiting forms of the potential energy that we have used). Now the coupling of D-particles to the matter on the brane can lead to a deformation of $\textsc{p}(\eta)$ on the brane which can lead to $\cal{E}$ having either sign. 

This would be expected to be a generic effect given such a $\textsc{p}(\eta)$, a demonstration of which can be seen in the context of a simple toy model: A D-particle is point-like from the point of view of observers on the brane. The scale of the strongest interactions is sub-stringy and so, in comparison, the size of the  bulk dimension is very large. The D-particle, as far as matter excitations of the string are concerned, has no quantum numbers. In order to gain some intuition about changes to $\textsc{p}(\eta)$ due to D-particle brane interactions we will consider a very simple  field theory where we represent the D-particle by a massive scalar field $\phi$ and the matter excitation by a harmonic oscillator on the boundary. Generically the model can be represented as an action:
\begin{equation}\label{toy}
    S=\frac{1}{2}\int_{\Omega}d^{d+1}\eta \sqrt{-G}(G^{MN}{\phi_{,M}}{\phi_{,N}}+m^{2}{\phi}^{2})
    \;-{\frac{1}{2}}\int_{\partial\Omega} d^{d}\eta \sqrt{-g} [g^{\mu\nu}q_{,\mu}q_{,\nu}+\mu^{2}q^{2}+2 \beta \phi q + \lambda  \phi^{2}]
\end{equation}
where $\Omega$ is the bulk, $\partial\Omega$ is the brane (which is the boundary of the bulk), $G^{MN}$ is the bulk metric, and $g^{\mu\nu}$ is the brane metric; the analysis is simpler for $d=0$ and suffices to illustrate the deformation of the distribution of $\textsc{p}(\eta)$ owing to bulk-brane interactions. The quantum theory is determined by the classical modes. For $d=0$, bulk $-\infty < \eta <0$, brane position $\eta=0$, conjugate momentum  $\pi(\eta,t)$, the action $S$ leads to a hamiltonian \cite{George}  
\begin{equation}\label{hamiltonian}
    H=\int_{-\infty}^{0}(\frac{1}{2}\pi(\eta,t)^{2}+\frac{1}{2}(\partial_{\eta}\phi(\eta,t))^{2}+\frac{1}{2}m^{2}\phi(\eta,t)^{2})dx
    +\frac{1}{2}\lambda \phi(0,t)^{2}+\beta \phi(0,t) q(t)+ \frac{1}{2} \mu^{2} q(t)^{2}+ \frac{1}{2}p(t)^{2}.
\end{equation}
The equation of motion is
\begin{equation}\label{classical}
    \partial_{t}^{2}\phi(\eta,t)= \partial_{\eta}^{2}\phi(\eta,t) -m^{2}\phi(\eta,t)-\delta(\eta)(\partial_{\eta}\phi(0,t)+\lambda \phi(0,t)+\beta q(t))
\end{equation}
which is equivalent to the solution of the massive Klein-Gordon equation with a boundary condition
\begin{equation}\label{boundary}
    \partial_{x}\phi(0,t)=-\lambda \phi(0,t)-\beta q(t).
\end{equation}
The oscillator satisfies the equation of motion
\begin{equation}\label{classical2}
    \partial_{t}^{2}q(t)=-\beta \phi(0,t)-\mu^{2}q(t).
\end{equation}

The mode solutions corresponding to (\ref{classical}) and (\ref{classical2}) are
\begin{equation}\label{mode}
    \phi(\eta,t)=\int_{0}^{\infty}(\rho(\rho^{2}-\mu^{2}+m^{2})\cos(\rho \eta)-(\lambda (\rho^{2}-\mu^{2}+m^{2})+\beta^{2)}\sin(\rho\eta))
    \times (a(\rho) \cos(\omega_{\rho}t)+b(\rho) \sin(\omega_{\rho}t)) d\rho
\end{equation}
and 
\begin{equation}\label{mode2}
    q(t)=\int_{0}^{\infty}\beta \rho (a(\rho)\cos(\omega_{\rho}t)+b(\rho) \sin(\omega_{\rho}t)) d\rho
\end{equation}
where $\omega_{\rho}=\sqrt{m^{2}+\rho^{2}} $ and $a(\rho)$ and $b(\rho)$ are real. The equations of motion are linear and so the integral over $\rho$ represents a linear superposition. These are solutions which propagate in the bulk. Square integrable states bound to the brane also exist by considering specific pure imaginary $\rho$. The asymptotic behaviour as $\eta\rightarrow {-\infty}$ has to be non-divergent and this selects the details of the coefficients in the solution. Once the solutions are substituted in $H$ we find a description in terms of harmonic oscillators associted with the classical modes.  Clearly from the mode structure for real $\rho$ (or bulk modes) we have an oscillatory behaviour with $\eta$ and so this will lead to a modulation in $\textsc{p}(\eta)$ as a result of (the analogue of) D-particle recoil and hence this will allow both positive and negative $\delta E$. The bound states do not participate in this modulation. A more detailed model involving two harmonic oscillators (with distinct frequencies) on the brane will be presented elsewhere but the qualitative features that we have presented here should not change. 

In the next section a string theoretic picture of flavour changes owing to non-BPS (compactified) D-particles will be considered. We shall argue in favour 
of the positivity of the energy density of the \emph{flavour-vacuum}~\cite{blasone}  and, hence, its appropriateness to represent the physical ground state of the D-foam in the presence of flavoured matter.

\subsection{D-foam-Induced Fermion Condensates and Flavour Vacuum Stability }

In the above formalism, the presence of mixed-sign fluctuations may lead to negative contributions to the vacuum energy, implying an \emph{instability} of the standard \emph{mass eigenstate vacuum} in the case of flavoured particles.~\footnote{We also note in passing, that in the case of anti-particles, viewed as ``negative energy'' solutions of the Free Dirac equation, ${\overline E}_{ < 0}=-\sqrt{p^2 + m_1^2}$, Eq.~(\ref{energycons}) implies $E_2 = - \overline{E}_{2,>0} = -\sqrt{p^2 + m_1^2} -  \frac{g_s}{2 M_s}\sigma_\parallel^2 \, p^2 + \delta E_{\rm short}(\sigma_{\rm short}'(p)) + \delta E_{\rm long}(\sigma_{\rm long}')$,
where, on adopting Feynman's interpretation, that anti-particles have positive energy but move ``backwards'' in time, we have denoted the positive energy of the flavoured antiparticle after interaction with the foam by ${\overline E}_{2, > 0}$.
Solving for ${\overline E}_{2, > 0}$, we thus obtain different modified dispersion relations between particle and antiparticle, due to the \emph{relative sign difference} of the foam corrections, and thus induced CPT violation. The difference in energy between particles and antiparticles, implied by the foam contributions, may lead to neutrino/anti-neutrino oscillations, which however we shall not explore here.} The qualitative description of the role of D-particles in producing an instability in the mass vacuum can be put on a more explicit footing. In order to do this we will need to have a more detailed picture of stringy matter and how the standard model of particle physics (and beyond) emerge. This represents a vast area known as string phenomenology where a multitude of detailed models involving stacks of  D-branes and orientifolds interacting with stringy matter have been constructed. In D-foam models certain generic four fermion  interactions are induced among fermionic matter which should arise in a large class of models. The interactions are of a four fermion vector-current-current form which can lead to a mass-matrix type mixing in the case of models where right-handed neutrinos exist. This type of four-fermion interaction describes the result of the splitting of open strings when interacting with the D-particle defects. Such interactions arise in D-foam models, based on type IA~\cite{emnnewuncert} ot type IIB strings~\cite{li}, as a consequence of intermediate open string exchanges, stretched between D-particles and the brane worlds; these represent quantum fluctuations of the recoiling D-particle during its interaction with the propagating matter open string states. Some of these exchanged states are U(1) gauge particles, where U(1) are Abelian gauge groups that are abundant in string models, not necessarily representing electromagnetism. As a result of such Abelian gauge particle exchanges, there will be induced, among others, attractive four fermion interactions in a low-energy effective field theory framework~\cite{antoniad}.
For chiral fermions, $\psi, \psi '$ for instance, such interactions (obtained by considering the corresponding tree-level (Veneziano-like) four-point string scattering amplitudes) assume the generic form~\cite{antoniad,li}
\begin{equation}\label{fourfermi}
\mathcal{L}_{\rm 4-eff } = \frac{\xi}{M_s^2} V^c \sum_{i,j} \sum_{a,b=L, R} \mathcal{G}_{a b} \overline{\psi_i}^{a} \gamma^\mu \psi_{i\,a}  \, \overline{\psi_j}^{\prime \,b} \gamma^\mu \psi^\prime_{j\,b}
\end{equation}
where  $V^c$ denotes the compactification volume element to four space-time dimensions, in units of the string length $\sqrt{\alpha'}$; the indices $i,j$ refer to fermion species, including flavour, $\mathcal{G}_{ab}$ are numerical coefficients depending on the (square of the) couplings of the particular interactions, and are of order $O(1)$ as far as the string scale $M_s$ is concerned, and $L,R$ denote the appropriate chirality of the spinors $\psi_i, \psi'_j$. The constant $\xi$ depends on the type of D-foam considered.
The combination
\begin{equation}\label{planckscale}
\kappa = 8\pi \frac{V^c}{M_s^2}
\end{equation}
defines the gravitational coupling of the four-dimensional effective theory.

For type IA string D-foam~\cite{Dfoam,emnnewuncert}, the D-particle defects, being point-like and electrically neutral,
interact only with electrically neutral particles, since charge conservation prevents the splitting (\emph{cf.} fig.~\ref{fig:recoil}) of charged open string states when interacting with the defect. In this case, the fermions $\psi$ represent neutrinos of the standard model, assuming
these to be the dominant electrically neutral fermions in the low-energy world today. Thus, $\xi=1$ for such neutral excitations, while $\xi =0$ for charged fermions.
On the other hand, as explained in \cite{li} and mentioned above, for D-foam models based on type IIB strings,  the D-``particles'' are only effectively point-like, being represented by D3 brane defects wrapped up along three cycles appropriately, which then intersect the D7 branes of the model (representing after compactification to three spatial dimensions our brane world). In such models, there are non-trivial charged particle interactions with the D-particles, of the type considered in fig.~\ref{fig:recoil}, which however are suppressed as compared to the neutral particle interactions by a factor:
\begin{equation}\label{eta}
\xi = \frac{(1.55)^4}{M_s^4 \, V_{A3} \, R'}
\end{equation}
where $M_s$ is the string scale, $V_{A3}$ denotes the volume of a cell on the D7 brane (playing the r\^ole of a brane Universe after compactification of four of its dimensions ) containing a D3 particle in the foam, and $R'$ is the fourth dimension of the D7 brane, transverse to the
D3-brane dimensions.  The corresponding four fermion interactions assume the form (\ref{fourfermi}) with $\xi$ given by (\ref{eta}).

Let us consider flavoured neutrinos which are chiral, where
\begin{equation}\label{chiral}
\gamma_5 \psi_i = \pm \psi_i~,
\end{equation}
and the $\pm$ sign depends on whether the fermion is left or right handed.
Indeed, in models such as the Standard Model we encounter only one type of Chiral fermion,
say left handed, satisfying (\ref{chiral}), with no right-handed neutrinos.
Next, let us  assume that there are \emph{right handed flavoured fermions} (e.g. neutrinos) interacting with the
extra U(1) stringy symmetries, which are different from the Abelian gauge group of the standard model interactions. In this sense, the presence of right handed fermions is consistent with their nature as being neutral under the standard model group.~\footnote{The assumptions on having right handed neutrinos in models beyond the standard model is not uncommon, see for instance \cite{righthanded}.} With the above in mind, we observe that, upon a Fierz rearrangement,
 the four-fermion interactions (\ref{fourfermi}),  involving both left- and right-handed flavoured fermions, includes terms :
\begin{equation}\label{fierz}
\mathcal{L}_{\rm 4-eff } \ni  \frac{\xi \mathcal{G}V^c}{M_s^2} \sum_{i,j}  \left[ -(\overline{\psi_i} \psi_j )\, (\overline{\psi_j} \psi_i)
+ (\overline{\psi_i} \gamma_5 \psi_j )\, (\overline{\psi_j} \gamma_5 \psi_i)+ \frac{1}{2} (\overline{\psi_i} \gamma^\mu \psi_{j} ) \, (\overline{\psi_j } \gamma^\mu \psi_{i}) +
\frac{1}{2} (\overline{\psi_i} \gamma_5 \gamma^\mu \psi_{j} ) \, (\overline{\psi_j } \gamma_5 \gamma^\mu \psi_{i})  \right]~,
\end{equation}
where for brevity and concreteness we have assumed one type of Non-Standard-Model interaction among neutrinos due to the intermediate string states, with $\mathcal{G} > 0$ (which is proportional to the square of the pertinent coupling), and
the spinors have both left and right-handed components, \emph{i.e.} $\psi_i = \begin{pmatrix} \psi_{L} \\ \psi_{R} \end{pmatrix}_i ~, \quad i = 1, \dots N$ where $N$ is the number of flavours. We can obtain \emph{attractive} interactions in the scalar channel of the form
\begin{equation}\label{fierz2}
\mathcal{L}_{\rm 4-eff } \ni  - \frac{\xi \mathcal{G}V^c}{M_s^2} \sum_{i,j}  \langle (\overline{\psi}_{i} \psi_j )\rangle  (\overline{\psi}^{j} \psi^i) ~.
\end{equation}
The condensates would lead to dynamical mass matrix contributions:
\begin{equation}\label{massmatrixdfoam}
m_{ij} = \frac{\xi \mathcal{G}V^c}{M_s^2}  \langle \overline{\psi}_i \psi_j \rangle, \quad i,j = 1, \dots N~.
\end{equation}
From the off-diagonal in flavour space parts of the condensate one would thus obtain
\emph{mixing}, while the diagonal parts would lead to
\emph{dynamically} generated \emph{fermion-mass} contributions, which can lead to chiral symmetry breaking.~\footnote{In the context of our D-brane Universes, the vacuum energy contributions due to the bulk D-particles (\emph{cf}. (\ref{D0-D8-long}), (\ref{pot1})) lead to (anti)de Sitter space-times.
Chiral symmetry breaking in such non-trivial space-time backgrounds has been discussed in the context of early Universe studies in 
\cite{candelas,neubert}, with the conclusion that the dynamical formation of chiral condensates, due to four-fermion interactions, exhibits 
better UltraViolet (UV) behaviour than in the corresponding flat space-time case~\cite{nambu}. We postpone a detailed discussion of such issues to a forthcoming publication.}
We stress once again that such scalar condensates characterize only theories where right-handed fermions do exist,
because otherwise
the scalar four-fermion terms
arising from the Fierz rearrangement (\ref{fierz}) are cancelled  by the pseudoscalar ones.

An important remark is in order here. The linearization of  the attractive four-fermion interactions in the scalar channel
by means of the non-propagating Hubbard-Stratonovich condensate field $\Delta_{ij}$, i.e.
\begin{equation}\label{fierz3}
\mathcal{L}_{\rm 4-eff } \ni  - \sum_{i,j} \left( \frac{M^2_s} {4 V^c | \xi | \mathcal{G}} \Delta_{ij} \Delta^{ij}  + \Delta_{ij}  (\overline{\psi_j} \psi_i)   \right)~
\end{equation}
would lead, upon formation of the condensate,  to \emph{negative} contributions to the \emph{vacuum energy} ($\mathcal{O}((\langle\Delta_{ij}\rangle)^2)$ terms above). However, in the D-foam situation, there are also the \emph{positive} energy contributions arising from the (temporal components of the) repulsive vector four-fermion condensates  in  (\ref{fierz}), which are non zero as a result of the flavour vacuum
that exists due to the induced mixing.  Indeed, a non-trivial vacuum condensate of the time component of the current operator $\langle \sum_{i} \overline{\psi}_i \gamma^0 \psi_i  \rangle = \langle \sum_{i} \psi^\dagger \psi_i \rangle $ is equivalent to a non-zero particle-species number, which is known to characterize the 
\emph{flavour vacuum}~\cite{blasone,mavrosarkar,mst1} but \emph{not} the mass-eigenstate one. 
These contributions could overcome the negative vacuum energy contributions on the brane worlds, arising from the interactions of the bulk D-particles with the brane Universe (\emph{cf.}  (\ref{pot1})), leading to the \emph{stability} of the flavour-vacuum. 

Incidentally, the fermion number condensates  \emph{violate} not only Lorentz  but also CPT symmetry. 
In the presence of such fermion-number condensates, the effective four-fermi interactions (\ref{fierz}), when linearized by means of appropriate Hubbard-Stratonovich fields, lead to 
a low-energy effective Lagrangian containing CPT and Lorentz violating terms of the type appearing in the parametrization of ref.~\cite{kostel}. This induced violation of Lorentz and CPT invariance is in agreement with the generic properties of the D-foam discussed previously.

\subsection{On the Scale of flavour-vacuum-condensate-induced Supersymmetry Breaking}

Before closing this section a final remark is in order about the nature  of symmetries that may be broken by the afore-mentioned flavour-vacuum condensates in D-foam models and the associated order of magnitude of the pertinent symmetry-breaking scales.
In addition to the chiral symmetries, that at any rate do not necessitate the introduction of a flavour vacuum, it is also supersymmetry that is broken in a novel non-perturbative way by the introduction of the flavour vacuum. In fact, the four-fermion interactions (\ref{fourfermi}), due to  intermediate-string-state exchanges,   arise consistently with target-space supersymmetries of the respective string theories~\cite{antoniad}, so formally the respective Lagrangians can be made supersymmetric, in the sense of their invariance under supersymmetry transformations (local supersymmetry (supergravity) actually, in view of the models' coupling to gravity).

However, the formation of Lorentz-violating fermion-flavour-vacuum condensates will induce a novel breaking of global supersymmetry, which we shall discuss in the next section (any extension to supergravity is left for future works).  It becomes obvious from the above discussion that the order of the induced supersymmetry breaking is then dictated by the order of the respective condensates.  In our stochastic foam situations, the
natural four-dimensional mass scale is $1/\sqrt{\kappa}$ (\ref{planckscale}), which from a four-dimensional point of view may be identified with the Planck scale, and for large radii compactification is different from the string scale $M_s$, which can be as low as a few TeV in the modern version of string theory. However, as discussed in \cite{mavrosarkar,mst1}, the induced condensates in D-foam models are calculated by subtracting appropriately the Minkowski vacuum contributions in the absence of D-particles. This implies~\cite{mavrosarkar,mst1} that the
condensates are \emph{proportional} to the average of stochastic fluctuations of the recoil velocities $\sigma_{\rm short, \, long}^{\prime \,2} \equiv {\sigma^{\prime \,2}}$
in (\ref{foam}), assuming for simplicity a common scale of D-Foam recoil velocity fluctuations.
Intuitively one can understand this dependence as follows: the flavour vacuum is Lorentz violating, and in D-Foam models the only source of local Lorentz violation are the stochastic fluctuations of the recoil velocities of the D-particles (\ref{foam}). Thus, it is natural that the flavour-vacuum condensate is proportional to the latter.
This makes the effective supersymmetry breaking mass scale in D-foam models, due to flavour vacuum, of order
\begin{equation}\label{effscale}
M^2 \sim \frac{1}{\kappa} \sigma^{\prime \,2}~.
\end{equation}
The parameter $\sigma^{\prime \,2}$ is an arbitrary parameter of the stringy foam model, although naturalness criteria may lead us to assume that it is more or less of order one (but this assumption is non binding).

To recapitulate, the above analysis implies that the open string states,
representing flavoured particles, when interacting with the D-foam,
may  \emph{induce flavour mixing}, of the order of the foam fluctuations.
The vacuum on the brane world has broken supersymmetry, due to the relative (fluctuation) motion between the brane world and the nearby D-particles, as well as \emph{broken Lorentz invariance}, due to the fluctuating recoil velocities of the D-particle space-time defects. These features imply that the appropriate Fock space vacuum for such configurations cannot be the ordinary Lorentz invariant, supersymmetric vacuum. We have argued in \cite{mst1} that the physically correct vacuum in this case is the \emph{flavour vacuum } proposed by Blasone and Vitiello~\cite{blasone} to describe generically theories with mixing, which is \emph{orthogonal} to the mass eigenstate vacuum. Moreover, the fact that there is \emph{particle creation} in our string case and in the flavour vacuum~\cite{blasone}, as mentioned above, constitutes an additional consistency check of the identification of the latter with the physical vacuum in our D-foam model. Indeed, because of this property, the Lorentz-violating flavour vacuum leads to Lorentz- and CPT- violating vector
condensates (temporal components) that contribute positively to the vacuum energy of the brane world, cancelling negative anti-de Sitter type contributions on the brane vacuum energy (\emph{cf}. (\ref{pot1})), and therefore restoring stability of the ground state. It was also argued in \cite{mst1} that the particle-creation feature of the flavour vacuum leads to a dynamical appearance of a high-momentum cut-off, which is set by the mass scale of the flavoured particles. The argument was based on the maximal amount of particle creation below the cutoff.

In the following section, therefore, we proceed  to study the properties of  supersymmetric fluids, using an interesting toy example, that of the Wess-Zumino supersymmetric theory. In the above context, this QFT should be viewed as an ``effective'' theory on our brane world punctured by the flavour-change inducing D-particle foam. We shall assume simple mixing in the mass matrix, which from the above microscopic point of view may be viewed as the result of four-fermion interactions in the presence of right-handed neutrinos.
We shall first formulate the theory, and then demonstrate some generic but quite non-trivial properties, with important differences between the boson and fermion excitations, thereby implying further non-trivial and novel manifestations of supersymmetry ``breaking''. As we shall see, in supersymmetric theories both flavoured bosons and fermions form condensates in the flavour vacuum, but the total vacuum energy turns out to be positive, and moreover, the equations of state between bosons and fermions turn out to be different.
These features constitute a novel, and purely non-perturbative, way of breaking supersymmetry. In what follows we shall demonstrate these properties in free theories, reserving an extension to interacting field theory in a future publication~\cite{new}.


\section{The Flavoured Wess-Zumino Model}\label{setup}

The WZ model is historically a prototype where the basic features of a supersymmetric theory
may be formulated in a concrete and quite instructive way. In our work, motivated by the considerations of the previous section, we shall be dealing with extensions of the simplest version of the free WZ model, in which the fields carry non-trivial flavour quantum number and  exhibit mixing.  In the next subsection, the model and its particle content, is described  in some detail. We should again stress that the flat-space supersymmetric WZ model is viewed here as the simplest prototype of an effective low-energy theory, describing the dynamics of excitations on a spatially flat brane world in the model of the previous section. The initial supersymmetry among the fields is inherited from the underlying supersymmetries of the string/brane theory; the
SUSY breaking is due to the relative brane motion and will be incorporated,  in this context, by viewing the ``flavour'' vacuum as the physical vacuum, describing correctly the associated mixing. It is understood that a more sophisticated local supersymmetry (supergravity)
effective lagrangian should be used in realistic constructions, but this will not affect qualitatively the important features of the flavour-vacuum revealed by the study of the toy WZ model.

\subsection{Supersymmetric Model}\label{wzmodel}
The Lagrangian of the model reads \cite{sohnius, farrill}:
\be\label{lag}
\lag =\lag_{\psi}+\lag_{S}+\lag_{P}
\ee
with
\bea
\lag_{\Psi}&=& \bar{\Psi}_f(i\partial\!\!\!/-M_f)\Psi_f^T\nn
\lag_{\mathcal S}&=& \partial_{\mu}\mathcal S_f \partial^{\mu}\mathcal S_f^T-\mathcal S_f M_b \mathcal S_f^T\nn
\lag_{\mathcal P}&=& \partial_{\mu}\mathcal P_f \partial^{\mu}\mathcal P_f^T-\mathcal P_f M_b \mathcal P_f^T
\eea
where  $\Psi_f\equiv(\psi_A,\psi_B)$, $\mathcal S_f\equiv(S_A,S_B)$, $\mathcal P_f\equiv(P_A,P_B)$, $M_f\equiv\begin{pmatrix}
 m_A& m_{AB} \\
m_{AB} & m_{B} \end{pmatrix}$, $M_b\equiv\begin{pmatrix}
 m_A^2& m^2_{AB} \\
m^2_{AB} & m^2_{B} \end{pmatrix}$.

The fields $\psi_\iota$, $S_\iota$ and $P_\iota$ (with $\iota=A,B$) are respectively a Majorana spinor,
a scalar and a pseudo-scalar.\footnote{The \textit{pseudo}-scalar nature of $P$ plays a r\^ole in the interactive WZ case, which we do not analyze here; in the current work, it can be regarded just as another scalar field. Nevertheless, to distinguish it from $S$, we will keep the nomenclature \emph{pseudo-scalar}.}
For our purposes, they are considered to be fields of definite \textit{flavour}, $A$ or $B$.
The cross terms of the Lagrangian (\ref{lag}), \textit{i.e.} terms involving products of fields of different flavours,  disappear when one expresses the model in terms of new fields, obtained by appropriate rotations of $\Psi_f$, $\mathcal S_f$ and $\mathcal P_f$:
\bea\label{lag2}
\lag_{\Psi}&=& \bar{\Psi}_m(i\partial\!\!\!/-M_1)\Psi_m^T\nn
\lag_{\mathcal S}&=& \partial_{\mu}\mathcal S_m \partial^{\mu}\mathcal S_m^T-\mathcal S_m M_2 \mathcal S_m^T\nn
\lag_{\mathcal P}&=& \partial_{\mu}\mathcal P_m \partial^{\mu}\mathcal P_m^T-\mathcal P_m M_2 \mathcal P_m^T
\eea
with $\Psi_m\equiv(\psi_1,\psi_2)=\Psi_f \Omega^T$, $\mathcal S_m\equiv(S_1,S_2)=\mathcal S_f \Omega^T$, $\mathcal P_m\equiv(P_1,P_2)=\mathcal S_f \Omega^T$,
$M_1\equiv\begin{pmatrix}
 m_1& 0 \\
0 & m_2 \end{pmatrix}$, $M_b\equiv\begin{pmatrix}
 m_1^2& 0 \\
0 & m^2_2 \end{pmatrix}$ and $\Omega \equiv\begin{pmatrix}
\ct & \st \\
-\st & \ct \end{pmatrix}$
where $m_1$, $m_2$ and $\theta$ are suitable functions of $m_{A}$, $m_{B}$ and $m_{AB}$.
The new Lagragian, written as (\ref{lag2}), is just the sum of free Lagrangians for the fields $\psi_i$, $S_i$ and $P_i$, all with the same mass $m_i$ ($i=1,2$).
Therefore we can decompose these fields as
\be\label{psimass}
\psi_i (x)=\sum_{r=1,2}\int{\frac{d^3k}{(2 \pi)^{3/2}} \left[  a^r_i(\vk) u^r_i (\vk)e^{-i\w_i(k)t}
						+ v_i^{r} (-\vk) a_i^{r\dagger}(-\vk)e^{i\w_i(k)t}\right]e^{i\vk \cdot \vx}}
\ee
\be\label{smass}
S_i(x)=\int{\frac{d^3k}{(2 \pi)^{3/2}}\frac{1}{\sqrt{2 \w_i(k)}} \left[  b_i(\vk)e^{-i\w_i(k)t}+ b_i^{\dagger}(-\vk) e^{i\w_i(k)t} \right] e^{i \vk \cdot \vx}  }
\ee
\be\label{pmass}
P_i(x)=\int{\frac{d^3k}{(2 \pi)^{3/2}}\frac{1}{\sqrt{2 \w_i(k)}} \left[  c_i(\vk)e^{-i\w_i(k)t}+ c_i^{\dagger}(-\vk) e^{i\w_i(k)t} \right] e^{i \vk \cdot \vx}  }
\ee
with $\{a^r_i(\vk),a^{s \dagger}_j(\vq)\}=\delta_{rs} [b_i(\vk),b^\dagger_j(\vq)]=\delta_{rs} [c_i(\vk),c^\dagger_j(\vq)]=\delta_{rs} \delta_{ij} \delta^3(\vk-\vq)$ and $\w_i\equiv\sqrt{\vk^2+m_i^2}$ with $i=1,2$.
From now on, we will refer to these fields as the fields \textit{in the mass representation}, to be contrasted with the ones \textit{in the flavour representation}, which are characterized by definite flavour quantum number.


\subsection{Flavour Mixing Formalism}\label{BVformalism}

Following the BV formalism \cite{blasone}, we introduce the operator $G_{\theta}(t)$, connecting fields between the flavour and mass representations as follows:
\bea
\phi_A(x)&=&G^{\dagger}_{\theta}(t) \phi_1 (x)G_{\theta}(t)\nn
\phi_B(x)&=&G^{\dagger}_{\theta}(t) \phi_2 (x)G_{\theta}(t)
\eea
with $\phi=\psi$, $S$, $P$.
Using Baker-Campbell-Hausdorff (BCH) formula, $e^Y X e^{-X}=Y+[X,Y]+\frac{1}{2!}[X,[X,Y]]+\frac{1}{3!}[X,[X,[X,Y]]]+...$, and the (anti-)commutation relations between fields, it can be shown that
\be
G_{\theta}(t)=e^{\theta \int d\vx \left(X_{12}(x)-X_{21}(x)\right)}
\ee
with
\be
X_{12}(x)\equiv \frac{1}{2}\psi^\dagger_1(x)\psi_2(x)+i \dot{S}_2(x) S_1(x)+i \dot{P}_2(x) P_1(x)
\ee

The \textit{flavour vacuum} is defined as\footnote{Although
the \textit{flavour vacuum} state depends on time (and therefore it would be more appropriate to write $|0(t)\rangle_f$), nevertheless, for the sake of simplicity, we will omit from now on the $t$ dependence in the notation. Remarkably, as we shall see below, the important
features of this state, of interest to us here, turn out to be \textit{time independent}.}
\be
\rfv\equiv G^{\dagger}_{\theta}(t)\rmv
\ee
where $\rmv$ is usual vacuum defined by $a^r_i(\vk)\rmv=b_i(\vk)\rmv=c_i(\vk)\rmv=0$ ($i=1,2$).

Using the operator $G_\theta (t)$ it is also possible to define new operators
\be
\left\{
\begin{array}{c}
a^r_A(\vk,t)\equiv G^{\dagger}_{\theta}(t) a^r_1(\vk)G_{\theta}(t)\\
a^r_B(\vk,t)\equiv G^{\dagger}_{\theta}(t) a^r_2(\vk)G_{\theta}(t)
\end{array}
\right.
\;\;\;\;
\left\{
\begin{array}{c}
b_A(\vk,t)\equiv G^{\dagger}_{\theta}(t) b_1(\vk) G_{\theta}(t)\\
b_B(\vk,t)\equiv G^{\dagger}_{\theta}(t) b_2(\vk) G_{\theta}(t)
\end{array}
\right.
\;\;\;\;
\left\{
\begin{array}{c}
c_A(\vk,t)\equiv G^{\dagger}_{\theta}(t) c_1(\vk) G_{\theta}(t)\\
c_B(\vk,t)\equiv G^{\dagger}_{\theta}(t) c_2(\vk) G_{\theta}(t)
\end{array}
\right.
\ee
Using again BCH formula, one can show that
\bea\label{masstoflavour}
a^r_A(\vk,t)&=&\ct a_1^r (\vk)-\st \sum_s \left( W^{rs}(\vk,t)a^s_2(\vk)+Y^{rs}(\vk,t)a^{s\dagger}_2(-\vk) \right)\nn
a^r_B(\vk,t)&=&\ct a_2^r(\vk)+\st \sum_s \left(  W^{sr*}(\vk,t) a_1^s(\vk)+Y^{sr}(-\vk,t)a_1^{s\dagger}(-\vk)  \right)
\eea
\bea
b_A(\vk,t)&=&\ct b_1(\vk) +\st \left( U^*(k,t) b_2(\vk)+V(k,t) b_2^\dagger (-\vk)\right)\nn
b_B(\vk,t)&=&\ct b_2(\vk) -\st \left( U(k,t) b_1(\vk)-V(k,t) b_1^\dagger (-\vk)\right)
\eea
\bea
c_A(\vk,t)&=&\ct c_1(\vk) +\st \left( U^*(k,t) c_2(\vk)+V(k,t) c_2^\dagger (-\vk)\right)\nn
c_B(\vk,t)&=&\ct c_2(\vk) -\st \left( U(k,t) c_1(\vk)-V(k,t) c_1^\dagger (-\vk)\right)
\eea
where
\bea\label{WY}
W^{rs}(\vk,t)&\equiv& \frac{1}{2}\left( u^{r\dagger}_2(\vk)u_1^s(\vk)+v_2^{s\dagger}(\vk)v_1^r(\vk) \right)e^{i(\w_1-\w_2)t}\nn
Y^{rs}(\vk,t)&\equiv& \frac{1}{2}\left( u^{r\dagger}_2(\vk)v_1^s(-\vk)+u_1^{s\dagger}(-\vk)v_2^r(\vk) \right)e^{i(\w_1+\w_2)t}
\eea
with $u^r_i(\vk)$ and $v^r_i(\vk)$ the spinors in (\ref{psimass})
\bea\label{UV}
U(k,t)&\equiv& \frac{1}{2}\left(\sqrt{\frac{\w_1}{\w_2}}+\sqrt{\frac{\w_2}{\w_1}}\right)e^{-i(\w_1-\w_2)t}\nn
V(k,t)&\equiv& \frac{1}{2}\left(\sqrt{\frac{\w_1}{\w_2}}-\sqrt{\frac{\w_2}{\w_1}}\right)e^{i(\w_1+\w_2)t}.
\eea
These operators satisfy the (anti-)commutation relations
\be\label{comrelflavour}
\{a^{r}_{\iota}(\vq,t),a^{s\dagger}_{\kappa}(\vp,t)\}=\delta_{rs}[b_\iota(\vq,t),b^\dagger_\kappa(\vp,t)]=
\delta_{rs}[c_\iota(\vq,t),c^\dagger_\kappa(\vp,t)]=\delta_{rs}\delta_{\iota \kappa}\delta^3(\vq-\vp)
\ee
(all others being zero and $\iota,\kappa=A,B$)
and it is also true that $a^r_\iota(\vk)\rfv=b_\iota(\vk)\rfv=c_\iota(\vk)\rfv=0$.
Therefore we are allowed to build a new Fock space out of these ingredients (operators+\textit{flavour vacuum}), in the usual way. This is actually a \textit{new} Fock space, and not just a different representation of the old one constructed by means of $a_1^\dagger$, $a^\dagger_2$ and $\rmv$, since all the new states are found orthogonal to the old ones in  the thermodynamic limit (see \cite{blasone} for an exhaustive discussion of flavour states,
and \textit{vacua}  within the BV formalism).
This physical inequivalence leads to interesting consequences. As we have said previously, one of those is the non-trivial violation of Lorentz symmetry, manifested through the non-zero vacuum expectation values of the stress-energy tensor, and also of SUSY in a particular way, as we now proceed to discuss.


\section{Flavour-Vacuum Induced SUSY Breaking}\label{features}

In order to study the properties of the \textit{flavour vacuum}, we will calculate the \textit{flavour} vacuum expectation value
of the stress-energy tensor, namely
\be\label{setvev}
\lfv T_{\mu\nu}(x) \rfv
\ee
The stress-energy tensor for the theory (\ref{lag2}) is written as
\be\label{set2}
T_{\mu\nu}(x)=\sum_{i=1,2}\left( 2\partial_{\left(\mu\right.}S_i(x)\partial_{\left.\nu\right)}S_i(x)+2\partial_{\left(\mu\right.}P_i(x)\partial_{\left.\nu\right)}P_i(x)
+i\bar{\psi}_i(x)\gamma_{\left(\mu\right.}\partial_{\left.\nu\right)}\psi_i(x)\right)
-\e_{\mu\nu}\lag
\ee
In case of a perfect relativistic fluid, as the \textit{flavour vacuum} condensate turns out to be,
the non-vanishing elements of this matrix, which are actually the ones on the diagonal, have a specific
physical interpretation: $T_{00}$ represents the energy density of the fluid and $T_{ii}$
the pressure in the three spatial directions.
In particular, we are interested in the equation of state, given by
\be
w=\frac{\lfv T_{ii}(x) \rfv}{\lfv T_{00}(x) \rfv}
\ee

An important remark is in order.
A straightforward computation of $\lfv T_{\mu\nu}(x) \rfv$ leads to a divergent result.
In previous works, when the fermionic and the bosonic cases were
treated separately, a \textit{finite} result was achieved in two steps: first, a normal ordering with respect to the usual
vacuum was defined
\be\label{normal}
\lfv : \mathcal O : \rfv \equiv \lfv  \mathcal O  \rfv-\lmv  \mathcal O  \rmv
\ee
then, a cutoff in the momentum space was introduced.
In the context of Supersymmetry, the first step, the normal ordering with respect to the usual vacuum, is not needed, since
\be
\lmv T_{\mu\nu}(x) \rmv=0.
\ee
A formal cutoff remains as a regulator of the theory,
a physical interpretation for which can be provided by
a microscopic theory (such as the brane model of \cite{mavrosarkar,mst1}, described briefly in section \ref{sec:dfoam} above).
The BV formalism is used to provide  an effective (Lorentz-violating) field theory in the low energy limit: the cutoff
appears dynamically in the theory, and is
interpreted as the high energy scale, below which the low-energy effective BV field theory is valid.


\subsection{Outline of Calculational Steps}\label{calculations}

In order to demonstrate the novel SUSY breaking induced by the flavour vacuum and describe the collective behaviour (equation of state) of the relativistic fluids under consideration, we need first to evaluate the quantity (\ref{setvev}), where the expectation value is taken with respect to the \textit{flavour vacuum} state.
To this end,
we first simplify the quantum operator structure of the various quantities involved, and then we deal with the remaining functions over the momenta.

Specifically, the stress-energy tensor (\ref{set2}) is first expressed in terms of the fields $\psi_i(x)$, $S_i(x)$ and $P_i(x)$,
in the \textit{mass representation}. On expressing the fields in terms of the mass-eigenstate creation and annihilation  (\textit{mass} ladder) operators  (\emph{cf.} 
(\ref{psimass}, \ref{smass}, \ref{pmass})), we can write
\bea\label{mset}
T_{\mu\nu}(x)\!\!\!&=&\!\!\! \sum_{i=1,2} \sum_{r,s} \int d\vp d\vq \left[ L_{\mu\nu}^{rs}(\vp,\vq,m_i) a_i^{r\dagger}(\vp)a_i^s(\vq)+																																						 M_{\mu\nu}^{rs}(\vp,\vq,m_i) a_i^{r\dagger}(\vp)a_i^{s\dagger}(\vq)\right.+\nn
								&&+\left. N_{\mu\nu}^{rs}(\vp,\vq,m_i) a_i^{r}(\vp)a_i^{s}(\vq)+K_{\mu\nu}^{rs}(\vp,\vq,m_i) a_i^r(\vp) a_i^{s\dagger}(\vq)  \right]
\eea
with $r,s=-1,0,1,2$, defining $a_i^{0}(\vk)\equiv b_i(\vk)$ and $a_i^{-1}(\vk)\equiv c_i(\vk)$, with $L$, $M$, $N$, $K$ suitable functions of the momenta.
In (\ref{mset}),
the bra $\lfv$ and  ket $\rfv$ act on the \textit{mass} ladder operators
\bea\label{msetvev}
\lfv T_{\mu\nu}(x)\rfv\!\!\!&=&\!\!\! \sum_{i=1,2} \sum_{r,s} \int d\vp d\vq \left[ L_{\mu\nu}^{rs}(\vp,\vq,m_i) \lfv a_i^{r\dagger}(\vp)a_i^s(\vq)\rfv\right.+\nn
								&&+M_{\mu\nu}^{rs}(\vp,\vq,m_i) \lfv a_i^{r\dagger}(\vp)a_i^{s\dagger}(\vq)\rfv+N_{\mu\nu}^{rs}(\vp,\vq,m_i) \lfv a_i^{r}(\vp)a_i^{s}(\vq)\rfv+\nn																 &&\left.+K_{\mu\nu}^{rs}(\vp,\vq,m_i) \lfv a_i^r(\vp) a_i^{s\dagger}(\vq) \rfv \right].
\eea
In order to simplify the expectation values of pairs of \textit{mass} operators, we use
the
formulae
	\bea
	a_1^r(\vk)&=&\ct a_A^r(\vk,t)+\st \sum_s \left(W^{rs}(\vk,t)a_B^s(\vk,t)+Y^{rs}(\vk,t) a_B^{s\dagger}(-\vk,t)\right)\nn
	a_2^r(\vk)&=&\ct a_B^r(\vk,t)-\st \sum_s \left(W^{sr*}(\vk,t)a_A^s(\vk,t)+Y^{sr}(-\vk,t) a_A^{s\dagger}(-\vk,t)\right)
	\eea
	\bea
	b_1(\vk)&=&\ct b_A(\vk,t)-\st\left(U^*(k,t)b_B(\vk,t)+V(k,t)a^\dagger_B(-\vk,t)\right)\nn
	b_2(\vk)&=&\ct b_B(\vk,t)+\st\left(U(k,t)b_A(\vk,t)-V(k,t)a^\dagger_A(-\vk,t)\right)
	\eea
	\bea
	c_1(\vk)&=&\ct c_A(\vk,t)-\st\left(U^*(k,t)c_B(\vk,t)+V(k,t)a^\dagger_B(-\vk,t)\right)\nn
	c_2(\vk)&=&\ct c_B(\vk,t)+\st\left(U(k,t)c_A(\vk,t)-V(k,t)a^\dagger_A(-\vk,t)\right)
	\eea
  which can be obtained from (\ref{masstoflavour}) ($W,Y,U,V$ being defined in (\ref{WY}, \ref{UV})),
the (anti-)commutation relations (\ref{comrelflavour}), as well as the
the property of the \textit{flavour vacuum}
	\be
	a^r_\iota(\vk,t)\rfv=b_\iota(\vk,t)\rfv=c_\iota(\vk,t)\rfv=0~.
	\ee
Being left with functions of the momenta, the last step
is to simplify and evaluate the integrals over $\vp$ and $\vq$.

We next proceed, therefore, to give the energy density and pressure of the fluid, obtained by following the above procedure, and discuss the associated
novel SUSY breaking mechanism induced by the flavour condensates in the BV formalism.

\subsection{Energy Density and Pressure}\label{results}

\begin{figure}[ht]
\centering
\includegraphics[width=8cm]{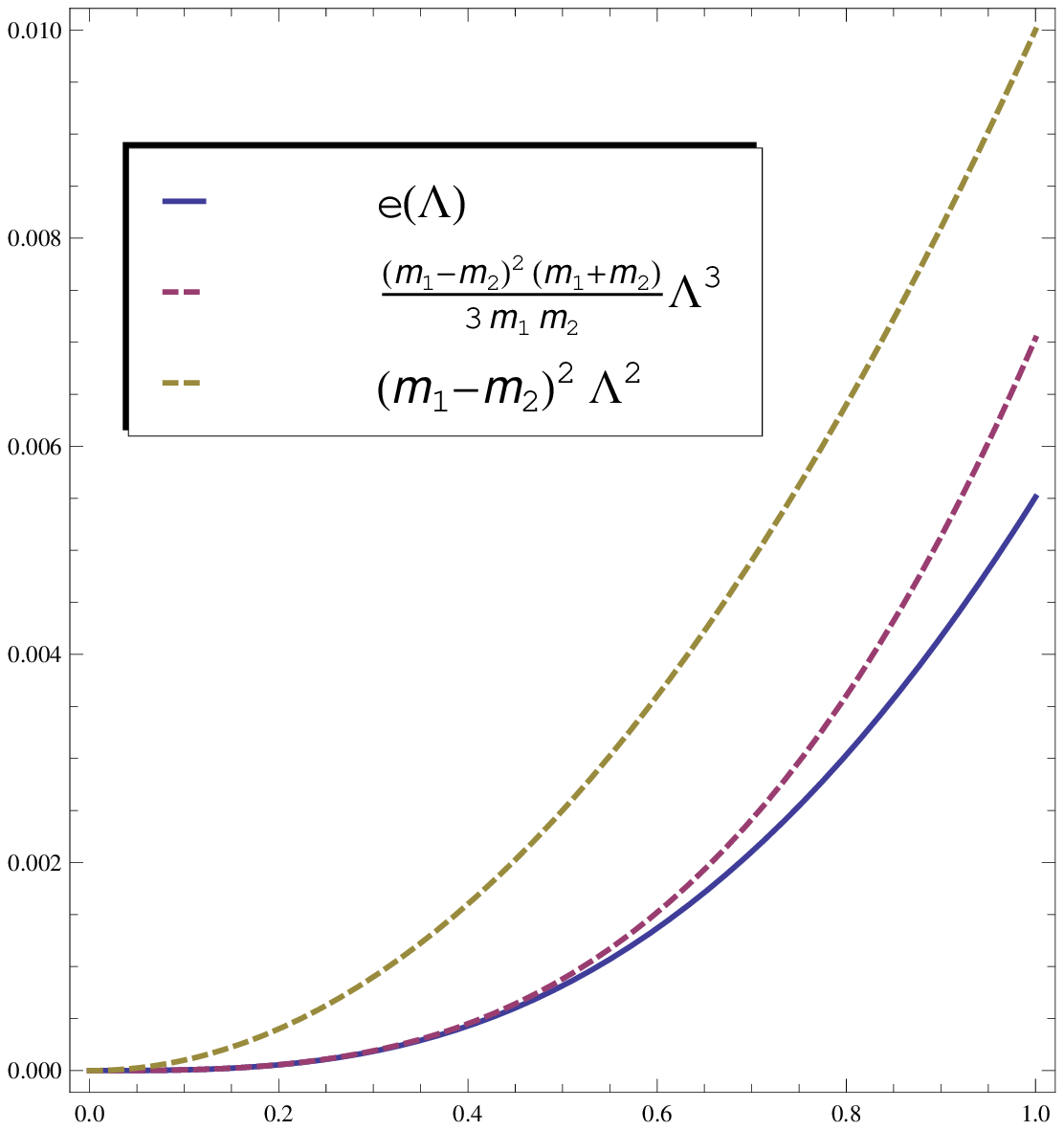} \hfill
\includegraphics[width=8cm]{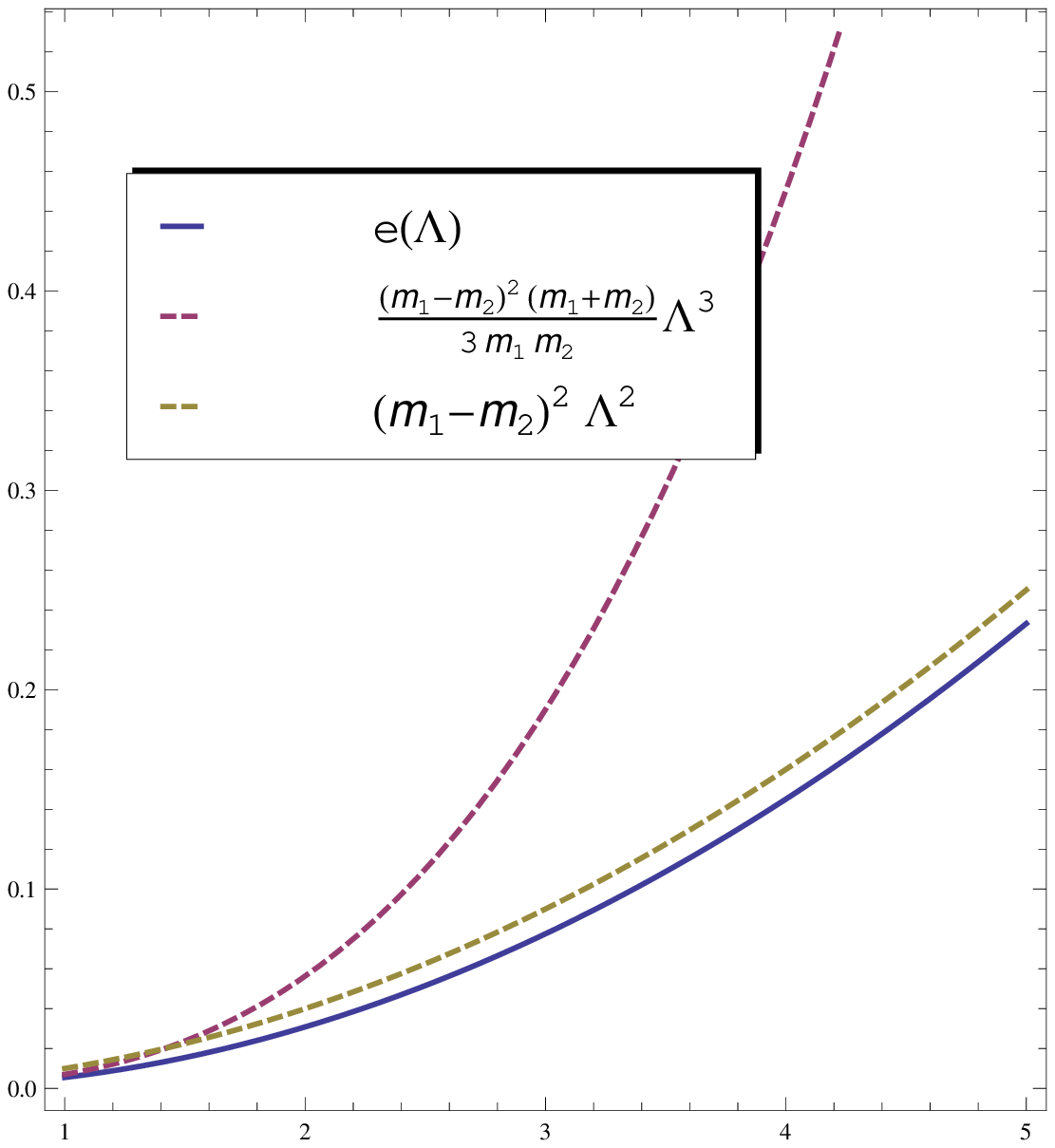}
\caption{The energy density as a function of the cut-off $\Lambda$, with the values: $m_1=1$, $m_2=0.9$,
in arbitrary units (sufficient for our purpose here, which is to demonstrate the dependence on $\Lambda$ only)
is plotted.
The two plots correspond to two different cut-off regimes:
the diagram on the left, shows the behaviour of $e(\Lambda)$ for small $\Lambda$. It approaches zero as $e(\Lambda)\sim \Lambda^3$ for $\Lambda \to 0$;
the plot on the right, depicts the behaviour of the energy density for large $\Lambda$, where we have $e(\Lambda)\sim \Lambda^2$. As a typical scale of the theory, with respect to which we define the low and high cut-off scale, we take
$m_1 \simeq m_2 = 1$. }
\label{fig:e}
\end{figure}

\begin{figure}[ht]
\centering
\includegraphics[width=8cm]{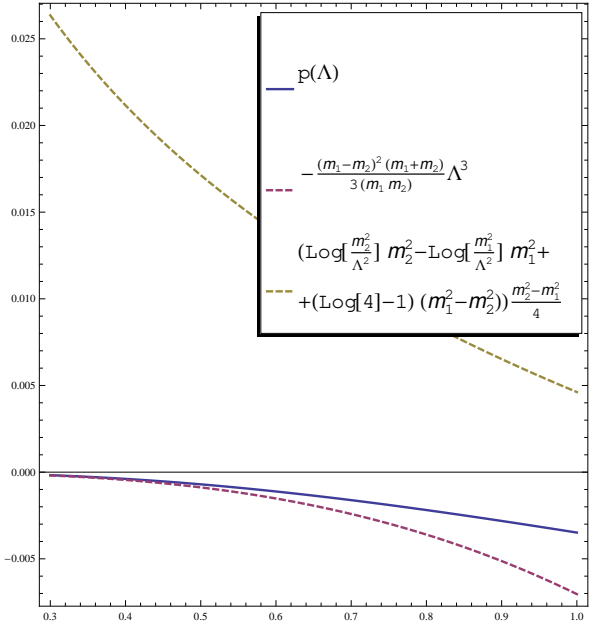} \hfill
\includegraphics[width=8cm]{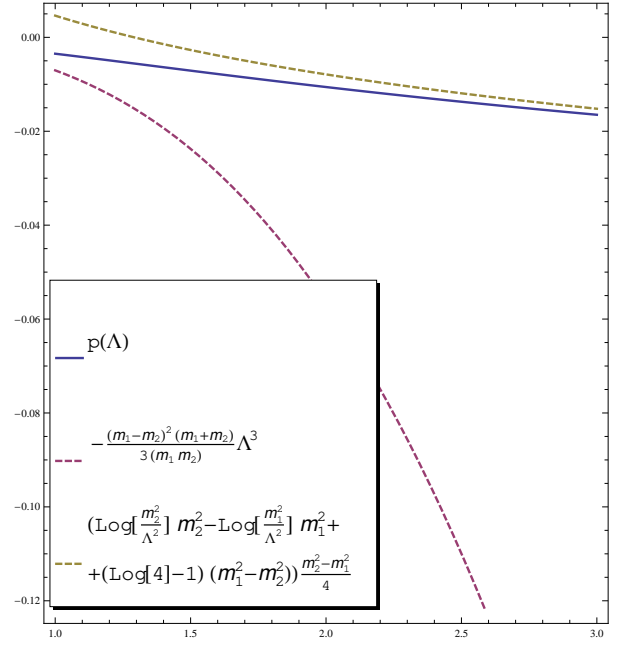}
\caption{The pressure as a function of the cut-off $\Lambda$, with the values: $\st=2 \pi^2$, $m_1=1$, $m_2=0.9$
is plotted.
Again, on the left we have the behaviour of $p(\Lambda)$ for small $\Lambda$ (compared with $m_2\approx m_1=1$),
where $p(\Lambda)\sim \Lambda^3$, and on the right for large $\Lambda$, where $p(\Lambda)\sim \log (\Lambda)$.}
\label{fig:p}
\end{figure}

After straightforward but quite involved calculations, one finds that the flavoured vacuum expectation value of the off-diagonal elements of the stress energy tensor \emph{vanish}, $_f\langle 0 | \,T_{\mu\nu} \,| 0 \rangle _f = 0$, $\mu \ne \nu$, as appropriate for a relativistic fluid (in a momentarily co-moving rest frame).
On the other hand, the energy density $e$ and pressure $p$ of the \textit{flavour vacuum} for the theory (\ref{lag}) are determined to be:
\bea\label{e1}
e&=&\lfv T_{00}(x)\rfv=\sst \frac{(m_1-m_2)^2}{2 \pi^2}\int_0^\Lambda dk\;k^2\left(\frac{1}{\w_1(k)}+\frac{1}{\w_2(k)}\right)=\nn
 &=&\sst \frac{(m_1-m_2)^2}{2 \pi^2}f(\Lambda)
\eea
\bea\label{p1}
p&=&\lfv T_{ii}(x)\rfv=-\sst \frac{m_1^2-m_2^2}{2 \pi^2}\int_0^\Lambda dk\; k^2 \left(\frac{1}{\w_1(k)}-\frac{1}{\w_2(k)}\right)\nn
 &=&\sst \frac{m_1^2-m_2^2}{2 \pi^2}g(\Lambda)
\eea
The integrals in (\ref{e1}) and (\ref{p1}) can be evaluated analytically. This yields for the functions $f(\Lambda)$ and $g(\Lambda)$:
\bea
f(\Lambda)&=&\frac{1}{2}\left( \Lambda \w_1(\Lambda)+\Lambda \w_2(\Lambda)-m_1^2 \log\left[\frac{\Lambda+\w_1(\Lambda)}{m_1}\right]
-m_2^2 \log\left[\frac{\Lambda+\w_2(\Lambda)}{m_2}\right]  \right) \nn
g(\Lambda)&=&\frac{1}{2}\left( \Lambda \w_1(\Lambda)-\Lambda \w_2(\Lambda)-m_1^2 \log\left[\frac{\Lambda+\w_1(\Lambda)}{m_1}\right]
+m_2^2 \log\left[\frac{\Lambda+\w_2(\Lambda)}{m_2}\right]  \right)\;\;\;\;\;\;
\eea
whereby $\Lambda$ we denote the cutoff in momentum space.
The behaviour of the energy density and pressure as functions of the cutoff is shown in figures \ref{fig:e} and \ref{fig:p}.

As we can see, the energy is \emph{positive}. This is not surprising, since we know that in a supersymmetric field-theory model the state with the lowest energy is unique, and corresponds to the usual vacuum $\rmv$, for which $\lmv T_{00} \rmv =0$,
all other states having positive energies. The difference in energy of the ``flavour vacuum'' from the mass-eigenstate one is due to the
condensation of massive flavoured particle states~\cite{blasone,capocosmo}.
The choice of this vacuum leads therefore to a \textit{breaking of supersymmetry}, in a novel way to be explored further below.

On the other hand, the pressure is \emph{negative}, leading to
an equation of state
\be
w=\frac{(m_1+m_2)}{(m_1-m_2)}\frac{g(\Lambda)}{f(\Lambda)}
\ee

\begin{figure}[ht]
\centering
\includegraphics[width=8cm]{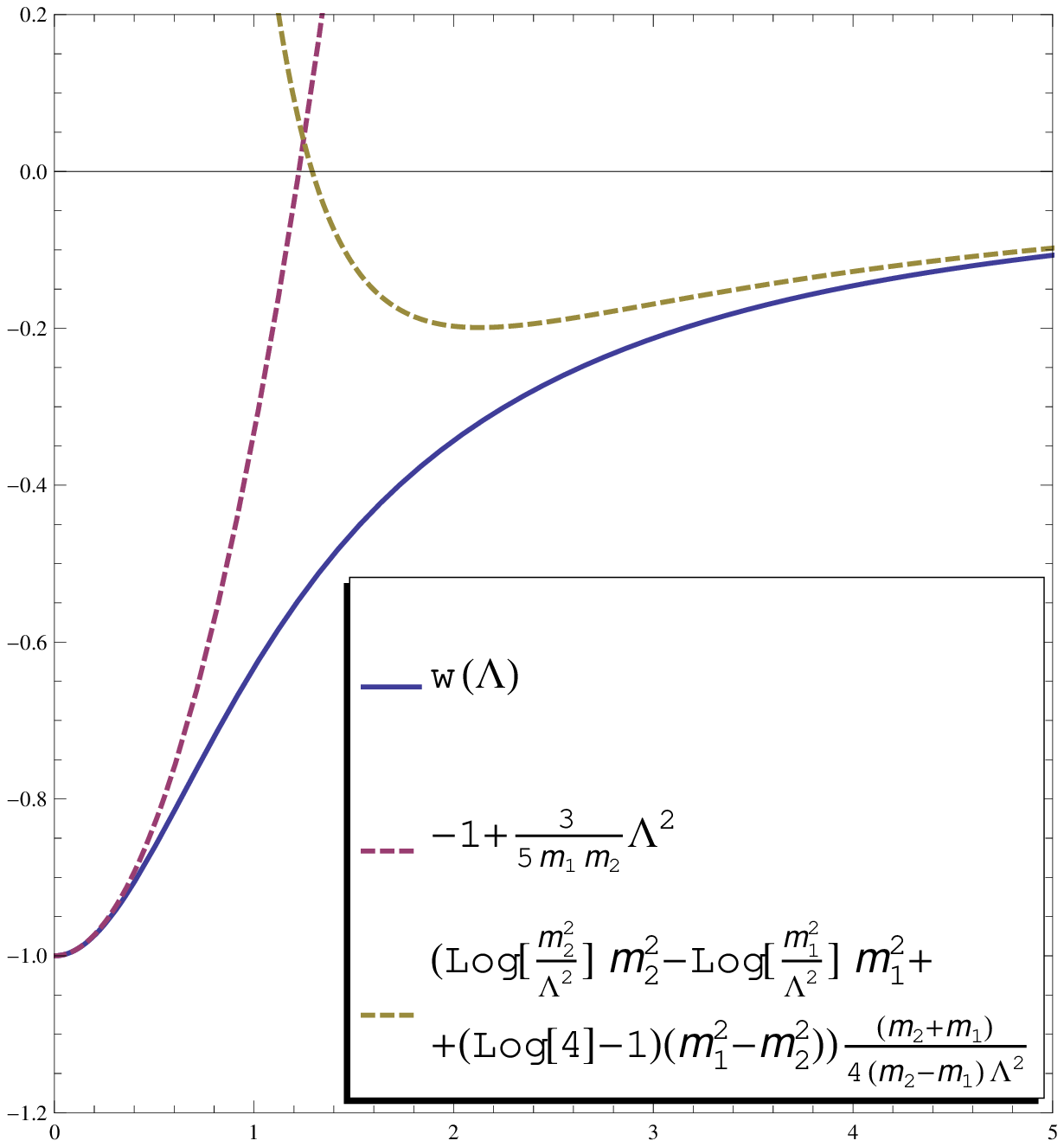}\hfill
\includegraphics[width=8cm]{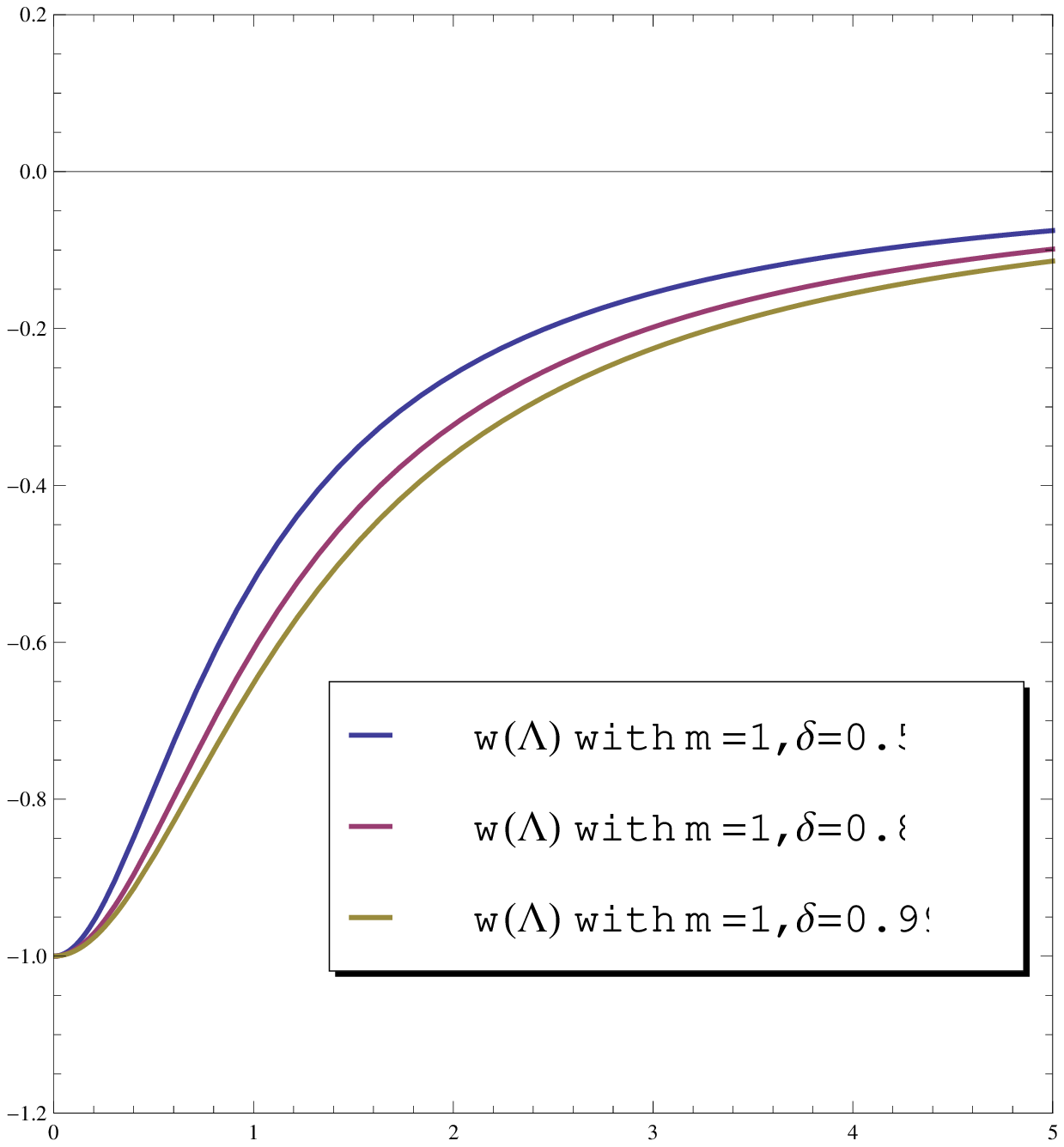}
\caption{The equation of state $w$ as a function of the cut-off is plotted;
The plot on the left corresponds to $m_1=1$, $m_2=0.9$. In the plot on the right we depict a set of three curves,
each of them  corresponding to different values of the mass-difference paramater $\delta$:
$m_2\equiv m + \delta$ and $m_1\equiv m$, with $m=1$ and $\delta=0.5,0.8,0.99$.}
\label{fig:w}
\end{figure}

As we can see from fig.~\ref{fig:w}, for small values of the cut-off (compared with $m_1\approx m_2$),
the equation of state approaches $-1$. This is the region in which our \textit{flavour vacuum} might give
a contribution to the Dark Energy, in the form of an approximate cosmological constant.
For greater values of the cutoff, the equation of state goes to zero as $-\log(\Lambda)\Lambda^{-2}M^2$,
with $M\equiv (m_1+m_2)^2/2$, and the fluid approaches that of a relativistic dust.~\footnote{The reader should recall at this stage that an accelerated expanding Universe is obtained when
$w<-1/3$, for the fluid that fills it in; the flavoured WZ fluid respects this property, for sufficiently low values of the cutoff $\Lambda$.
However, in the current work, we did not perform a self-consistent analysis of the flavour vacuum in such curved space-times, incorporating particle production, as done in \cite{mavrosarkar,mst1}. Thus, the conclusions one can reach from the above flat space-time analysis is that this type of fluids can potentially find applications to cosmology, once the above step is performed.}

\subsection{Different Equation of State for Bosons and Fermions - Strong \& Novel SUSY Breaking \label{sec:eos}}

In order to understand the above results and compare them with our previous work \cite{mavrosarkar, mst1},
it is useful to disentangle the various contributions
to $e$ and $p$ coming from the boson and Majorana fermion fields.

Since the Lagrangian (\ref{lag}) is the sum of six free lagrangians (two scalars,
two pseudo-scalars and two Majorana spinors), the stress-energy tensor can be written as the sum of three different terms:
\be
T_{\mu\nu}(x)=T_{\mu\nu}^S(x)+T_{\mu\nu}^P(x)+T_{\mu\nu}^\psi(x).
\ee
where the various superscripts denote scalar ($S$), pseudoscalar ($P$) and fermion ($\psi$) contributions, respectively.
The \emph{flavour vacuum} expectation values
$\lfv T_{\mu\nu}^S(x) \rfv~, \, \lfv T_{\mu\nu}^P(x) \rfv $ and
$ \lfv T_{\mu\nu}^\psi(x) \rfv $
are non-vanishing only for the diagonal elements.

As far as the energy density, $e$, is concerned, we have
\bea
\lfv \sum_{i=1,2}T_{00}^{S_i}(x)\rfv\!\!\!&=&\!\!\!\lfv \sum_{i=1,2}T_{00}^{P_i}(x)\rfv=\nn
																		\!\!\!&=&\!\!\!\int\!\!dk \frac{k^2}{2 \pi^2}\left(\w_1(k)+\w_2(k)\right)
																		\left[1+ \sst \frac{\left(\w_1(k)-\w_2(k)\right)^2}{2\w_1(k) \w_2(k)}\right]\\
\lfv \sum_{i=1,2}T_{00}^{\psi_i}(x)\rfv\!\!\!&=&\!\!\!\int\!\!dk \frac{k^2}{\pi^2}\left(\w_1(k)+\w_2(k)\right)
					\left[-1+ \sst\left( \frac{\w_1(k)\w_2(k)-m_1 m_2-k^2}{\w_1(k)\w_2(k)}\right)\right]\;\;\;\;\;\;\;\;
\eea
On the other hand, the respective contributions to the pressure are
\bea
\lfv \sum_{i=1,2}T_{ii}^{S_i}(x)\rfv\!\!\!&=&\!\!\!\lfv \sum_{i=1,2}T_{ii}^{P_i}(x)\rfv
					=\nn
					\!\!\!&=&\!\!\!\int\!\!dk \frac{k^2}{2\pi^2}\left[\frac{k^2}{3}\left(\frac{1}{\w_1(k)}+\frac{1}{\w_2(k)}\right)
					 -\sst \frac{m_1^2-m_2^2}{2}\left(\frac{1}{\w_1(k)}-\frac{1}{\w_2(k)}\right)\right]\;\;\;\;\;\;\\
\lfv \sum_{i=1,2}T_{ii}^{\psi_i}(x)\rfv\!\!\!&=&\!\!\!-\int\!\!dk\frac{k^4}{3 \pi^2}\left(\frac{1}{\w_1(k)}+\frac{1}{\w_2(k)}\right)
\eea
As we anticipated in the beginning of this section, such a theory needs  normal ordering with respect to the usual vacuum (\ref{normal}).
Now let us look at the individual equations of state:
\be
w_b=\frac{\lfv: \sum_{i=1,2}T_{ii}^{S_i}(x):\rfv}{\lfv: \sum_{i=1,2}T_{00}^{S_i}(x):\rfv}
					=\frac{\lfv: \sum_{i=1,2}T_{ii}^{P_i}(x):\rfv}{\lfv: \sum_{i=1,2}T_{00}^{P_i}(x):\rfv}=-1
\ee
and
\be
w_f=\frac{\lfv: \sum_{i=1,2}T_{ii}^{\psi_i}(x):\rfv}{\lfv: \sum_{i=1,2}T_{00}^{\psi_i}(x):\rfv}=0~.
\ee
This means that the bosonic contribution to the vacuum condensate
by itself would lead to a \emph{pure cosmological constant-type} equation of state~\cite{mavrosarkar,mst1}.
This behaviour of the fluid is mitigated by the contribution of the Majorana fermionic fluid, that has a \emph{vanishing} pressure.

When just low energies (induced by a low cutoff scale) are taken into account, the bosonic contribution drives the equation of state towards  $-1$, but as the cut-off increases, the fermionic contribution gets stronger and stronger, leading $w$ asymptotically to zero value, thus making the fluid behaving as dust in such high energy regimes.
The different behaviour between fermions and bosons indicates, in addition to the non-zero vacuum energy, a novel strong type of \emph{SUSY breaking} in the model, purely non-perturbative, induced by the various particle excitations of the flavour vacuum.

The choice of the cut-off is therefore crucial in determining
the behaviour of the fluid~\cite{capocosmo}. This implies the necessity for constructing models in which the cut-off scale is determined dynamically by the microscopic theory, as, for instance, the D-brane model of \cite{mavrosarkar,mst1}, a feature that cannot be captured by the simplified free WZ model considered here.

However, we should stress that the cut-off dependence of the condensate is a consequence of the fact that the computations have been done here in flat space-time, and thus some regularization is in order (the reader should compare, for instance, the cut-off dependence of the effective running coupling of the non-renormalizable four-fermion interaction
in the chiral-symmetry breaking model of ref.~\cite{nambu}, which enters the expression for the respective condensates.
When the analysis of chiral symmetry breaking is performed in curved, and in particular de Sitter space time~\cite{neubert}, as appropriate for
our D-foam framework, any potential Ultra Violet (UV)  infinities may be absorbed in the cosmological constant and Planck scale of the de Sitter space-time, as demonstrated explicitly in the related analysis of refs. \cite{candelas,neubert}. In our context one should expect a similar behaviour, since the flavour-vacuum condensates are also induced in our D-foam model from appropriate (non-renormalizable) four-fermion interactions  in the effective low-energy theory (\ref{fourfermi}), which become better behaved in the UV in our (anti) de Sitter backgrounds.  Hence the
presence of a flavour vacuum condensate is not an artefact of the UV cutoff.

\section{Discussion}

In this work we have examined a novel type of SUSY breaking in a flavoured supersymmetric WZ smodel, induced by  mixing in the BV formalism. The strong breaking of SUSY is characterized not only by a non-vanishing flavour-vacuum
energy but also by different equations of state between bosonic and fermionic particles. We have motivated the choice of the flavour vacuum over the mass eigenstate vacuum by an explicit, and by \emph{no means} generic, example of a space-time quantum foam in string theory.
The vacuum of such a configuration broke target-space supersymmetry and Lorentz symmetry, as a result of the quantum fluctuations of massive space-time defects. As such, the ordinary supersymmetric and Lorentz invariant vacuum is not appropriate for the description of the underlying physics, and the \emph{flavour vacuum} is a \emph{necessity}.
Despite the difference in the set-up from those in the models previously studied \cite{mavrosarkar,mst1}, SUSY breaking is
a feature also in this context.
Indeed, compared with \cite{mavrosarkar,mst1}, a flat universe instead of a FRW universe is assumed in the present work.
Moreover, no MSW effect has been introduced.
As a result, BV formalism by itself provides a mechanism for SUSY breaking.

Such a mechanism is not a \textit{spontaneous} SUSY breaking in the usual sense: no potential is introduced and formally, if the model is not considered as an effective theory,  a state with zero energy
(denoted by $\rmv$), exists. However, such a state does not represent the \textit{physical} vacuum; in the sense of \cite{blasone}
only flavoured particles can be produced/annihilated, if one does not insist on energy conservation at an interaction vertex (taking into account the time-energy uncertainty relation). From our D-particle-foam view point, energy is of course conserved overall in a process where a flavoured particle is produced  by the ``vacuum''. The missing energy from a low-energy field theory point of view~\cite{blasone}, amounting to the difference in energy between the normal mass-eigenstate and flavour vacua, is accounted for in the string-theory picture by the recoil of the D-particle defects in the foam and the induced condensation of particle-antiparticle flavoured particles. The latter amount to stretched string states between the recoiling D-particle and the D3-brane world in Fig.~\ref{fig:recoil}. Such \emph{back-reaction} effects of quantum flavoured matter onto the surrounding space-time involve stringy/brany degrees of freedom that go beyond the local field theory formalism.

The strong SUSY breaking that characterize such a state is a really new feature here.
It would be interesting to discuss other, phenomenologically more realistic models, involving Dirac fermions and, in general, interacting fields~\cite{new}. This would allow us to check whether the Majorana fermion fluid zero pressure we have found here is a specific feature of the free WZ model or is valid in general and study the patterns of flavour-induced symmetry breaking through the associated equations of state. One is tempted to  interpret the zero total pressure of the fermionic fluid as a result of an extra contribution, as compared with the bosonic case, due to the Pauli exclusion principle, the so called degeneracy pressure.
The latter is a positive contribution to the total pressure of a fermion fluid which is due to the extra force one has to exert as
a consequence of the fact  that two identical fermions cannot occupy the same quantum state at the same time. The force provided by this pressure sets a limit on the extent to which matter can be squeezed together without  collapsing into, \emph{e.g}. a neutron star or black hole. When two fermions (of the same flavour) are squeezed too close together, the exclusion principle requires them to have opposite spins in order to occupy the same energy level. To add another fermion of this flavour to a given volume (as required by the formation of a condensate) requires raising the fermion's energy level, and this requirement for energy to compress the material appears as a (positive) pressure. In our cosmological situation, it may be that this positive contribution to the pressure cancels out the cosmological negative pressure of the fermionic fluid, leading to the dust-like behaviour we find here. On the other hand, the bosons are not characterised by such degeneracy pressure contributions, and therefore their total (negative) pressure leads to an equation of state resembling that of a cosmological constant~\cite{mavrosarkar,mst1}. These issues, together with the extension of the formalism to incorporate interacting field theories,  are currently under investigation.


\section*{Acknowledgements}

The work of N.E.M. and S.S. is partially supported by the European
Union through the Marie Curie Research and Training Network \emph{UniverseNet}
(MRTN-2006-035863), while that of W.T. by a King's College London
postgraduate scholarship.


\newpage
\eject

\newpage

\eject
\newpage
\eject

\newpage


\begin{thebibliography}{99}

\bibitem{mavrosarkar} N.~E.~Mavromatos and S.~Sarkar,
  New J.\ Phys.\  {\bf 10}, 073009 (2008)
  [arXiv:0710.4541 [hep-th]].

\bibitem{mst1} N.~E.~Mavromatos, S.~Sarkar and W.~Tarantino,
  Phys.\ Rev.\  D {\bf 80} (2009) 084046
  [arXiv:0907.5122 [hep-th]].

\bibitem{blasone}
  M.~Blasone and G.~Vitiello,
  Annals Phys.\  {\bf 244} (1995) 283
  [Erratum-ibid.\  {\bf 249} (1996) 363]
  [arXiv:hep-ph/9501263].



	
\bibitem{birrell}
	N.~D.~Birrell and P.~C.~W.~Davies,
	\emph{Quantum Fields in Curved Space} (Cambridge University Press (1982)).



\bibitem{capocosmo} A.~Capolupo, S.~Capozziello and G.~Vitiello,
  Phys.\ Lett.\  A {\bf 363}, 53 (2007)
  [arXiv:astro-ph/0602467];
  Phys.\ Lett.\  A {\bf 373}, 601 (2009)
  [arXiv:0809.0085 [hep-th]], and references therein.

\bibitem{capolupo} A.~Capolupo, M.~Di Mauro and A.~Iorio,
  arXiv:1009.5041 [hep-th].



\bibitem{Dfoam} J.~R.~Ellis, N.~E.~Mavromatos and M.~Westmuckett,
Phys.\ Rev.\ D \textbf{70}, 044036 (2004) [arXiv:gr-qc/0405066];
\emph{ibid.} {\bf 71}, 106006 (2005)~.

 \bibitem{new} N.E. Mavromatos, Sarben Sarkar and W. Tarantino, in progress.


\bibitem{polchinski} See for instance: J. Polchinski, \emph{String Theory},
Vol. \textbf{2} (Cambridge University Press, 1998).



\bibitem{emnnewuncert} J.~R.~Ellis, N.~E.~Mavromatos and D.~V.~Nanopoulos,
  Phys.\ Lett.\  B {\bf 665}, 412 (2008)
  [arXiv:0804.3566 [hep-th]];


  \bibitem{li} T.~Li, N.~E.~Mavromatos, D.~V.~Nanopoulos and D.~Xie,
  Phys.\ Lett.\  B {\bf 679}, 407 (2009)
  [arXiv:0903.1303 [hep-th]].

\bibitem{sarkar} N.~E.~Mavromatos and S.~Sarkar,
  Phys.\ Rev.\  D {\bf 72}, 065016 (2005)
  [arXiv:hep-th/0506242].
  
 \bibitem{George}
  A.~George,
  J.\ Phys.\ A  {\bf 38} (2005) 7399
  [arXiv:hep-th/0412067].
 

\bibitem{antoniad}
I.~Antoniadis, K.~Benakli and A.~Laugier,
  JHEP {\bf 0105}, 044 (2001)
  [arXiv:hep-th/0011281];
D.~Lust, O.~Schlotterer, S.~Stieberger and T.~R.~Taylor,
  Nucl.\ Phys.\  B {\bf 828}, 139 (2010)
  [arXiv:0908.0409 [hep-th]].
D.~Lust, S.~Stieberger and T.~R.~Taylor,
  Nucl.\ Phys.\  B {\bf 808}, 1 (2009)
  [arXiv:0807.3333 [hep-th]].



  
    


 \bibitem{righthanded} There are many model employing right-handed neutrinos, ranging from purely phenomenological ones to
 microscopic models obtained from extensions of the standard model and string theory. These models contain four-neutrino interactions
 that may lead to condensates, with effects in the standard model physics and/or cosmology. For a very partial (and by no means complete) list of recent works, see:  G.~Barenboim and J.~Rasero,
  JHEP {\bf 1103}, 097 (2011)
  [arXiv:1009.3024 [hep-ph]], where cosmological consequences of a phenomenologically introduced four-right-handed-neutrino
  interactions have been explored.
  A.~Aparici, A.~Santamaria and J.~Wudka,
  J.\ Phys.\ G {\bf 37}, 075012 (2010)
  [arXiv:0911.4103 [hep-ph]];
  F.~del Aguila, J.~de Blas, R.~Szafron, J.~Wudka and M.~Zralek,
  Phys.\ Lett.\  B {\bf 683}, 282 (2010)
  [arXiv:0911.3158 [hep-ph]];
R.~Tatar, Y.~Tsuchiya and T.~Watari,
  Nucl.\ Phys.\  B {\bf 823}, 1 (2009)
  [arXiv:0905.2289 [hep-th]];
M.~Endo and T.~Shindou,
  JHEP {\bf 0909}, 037 (2009)
  [arXiv:0903.1813 [hep-ph]];
 S.~Antusch, J.~Kersten, M.~Lindner and M.~Ratz,
  Nucl.\ Phys.\  B {\bf 658}, 203 (2003)
  [arXiv:hep-ph/0211385].


\bibitem{candelas} P.~Candelas and D.~J.~Raine,
  Phys.\ Rev.\  D {\bf 12}, 965 (1975).
  
  \bibitem{neubert} F.~Giacosa, R.~Hofmann and M.~Neubert,
  JHEP {\bf 0802}, 077 (2008)
  [arXiv:0801.0197 [hep-th]];
  S.~Alexander, T.~Biswas and G.~Calcagni,
  Phys.\ Rev.\  D {\bf 81}, 043511 (2010)
  [Erratum-ibid.\  D {\bf 81}, 069902 (2010)]
  [arXiv:0906.5161 [astro-ph.CO]].

\bibitem{nambu} Y.~Nambu and G.~Jona-Lasinio,
  Phys.\ Rev.\  {\bf 122}, 345 (1961);
  \emph{ibid}.  {\bf 124}, 246 (1961).


\bibitem{kostel}  V.~A.~Kostelecky and M.~Mewes,
  Phys.\ Rev.\  D {\bf 70}, 076002 (2004)
  [arXiv:hep-ph/0406255]. See also: S.~H.~S.~Alexander,
  arXiv:0911.5156 [hep-ph].

 
  \bibitem{farrill}
  J.~M.~Figueroa-O'Farrill,
  arXiv:hep-th/0109172.

\bibitem{sohnius}
  M.~F.~Sohnius,
  Phys.\ Rept.\  {\bf 128} (1985) 39.




\end{thebibliography}
\end{document}